\documentclass[
  notitlepage,
  twocolumn,
  preprintnumbers,
  longbibliography,
  superscriptaddress,
  aps,
  prl,
  amsmath,
  amssymb,
]{revtex4-2}
\usepackage{listings}
\usepackage[dvipsnames]{xcolor}
\usepackage{graphicx}
\usepackage{multirow}

\usepackage{physics}
\usepackage{ifthen}
\usepackage{braket}
\usepackage[normalem]{ulem}

\usepackage{amsmath}
\usepackage{amsthm, amssymb}
\usepackage{slashed}
\usepackage{bm}
\usepackage[
  pdfpagelabels,
  plainpages=false,
  bookmarks=true,
  colorlinks=true,
  linkcolor=red,
  urlcolor=blue,
  citecolor=blue
]{hyperref}
\usepackage{dsfont}
\usepackage{changepage}
\usepackage{array}
\usepackage{bbm}
\usepackage[cal=boondox]{mathalpha}

\renewcommand{\vec}[1]{\bm{\mathbf{#1}}}

\newcommand{\loc}{\text{loc}}

\newcommand{\mm}[4]{\begin{bmatrix} #1 & #2 \\ #3 & #4 \end{bmatrix}}

\begin{document}
\title{Efficient simulation of long-range systems: ground state and real-time dynamics}

\author{Jia-Lin Chen}
\affiliation{Institute of Physics, Chinese Academy of Sciences, Beijing 100190, China}
\affiliation{School of Physical Sciences, University of Chinese Academy of Sciences, Beijing 100049, China.}

\author{Tao Xiang}
\email{txiang@iphy.ac.cn}
\affiliation{Institute of Physics, Chinese Academy of Sciences, Beijing 100190, China}
\affiliation{School of Physical Sciences, University of Chinese Academy of Sciences, Beijing 100049, China.}

\author{Yantao Wu}
\email{yantaow@iphy.ac.cn}
\affiliation{Institute of Physics, Chinese Academy of Sciences, Beijing 100190, China}

\date{\today}

\begin{abstract}
Long-range interactions are ubiquitous in quantum systems, but are among the least accessible for numerical many-body methods.
In variational Monte Carlo (VMC) calculations of $N$-site quantum spin systems with arbitrary all-to-all two-body couplings, evaluating the local energy generally costs $O(N^3)$. 
We introduce an adaptive-basis framework that reduces this cost to $O(N)$ for tensor network states, enabling scalable and accurate calculations of ground states and real-time dynamics, substantially expanding the scope of classically simulable long-range quantum systems.
As a result, we address the fundamental question of ordering in the ground states of the antiferromagnetic $1/r^\alpha$ XY model with varying $\alpha$, which was previously inaccessible at size $20\times20$. 
We find that in-plane antiferromagnetic order persists for $\alpha \ge 2$ but is not observed at $\alpha = 1$ up to $20\times20$.
As a further application, we simulate the adiabatic real-time protocol of a $10\times10$ dipolar XY model realized in a Rydberg simulator [C. Chen et al., Nature 616, 691 (2023)], which was previously beyond the reach of controlled classical simulation.
Our work establishes tensor network VMC both as a powerful numerical approach to two-dimensional long-range quantum spin systems and as an effective simulator for long-range quantum platforms, which are currently in urgent need of scalable classical benchmarking tools. 
\end{abstract}

\maketitle
\addtocontents{toc}{\protect\SMtocoff}

\textit{Introduction.}
Long-range interactions appear naturally in quantum materials, arising from dipoles in magnetic atoms~\cite{dePaz2013}, polar molecules~\cite{Gorshkov2011,Yan2013,Christakis2023}, and solid-state spin ensembles~\cite{Zu2021}, and qualitatively change many-body physics \cite{Defenu2023}.
They are also the native setting of analog quantum simulators.
Rydberg-atom arrays~\cite{SaffmanWalkerMolmer2010,BrowaeysLahaye2020} and trapped ions~\cite{PorrasCirac2004,Kim2009,Monroe2021} now realize two-dimensional long-range spin models with hundreds of particles~\cite{Britton2012,Bohnet2016,Chen2023,Bornet2023}, and operate through real-time protocols.
At these sizes, controlled classical benchmarks are urgently needed, especially in the regime where these devices operate: two dimensions, long-range interaction, and real time.

This regime is closed to existing numerical methods.
Quantum Monte Carlo (QMC) handles unfrustrated couplings in equilibrium~\cite{Sandvik2003}, but has no access to real time and meets a sign problem~\cite{TroyerWiese2005} in the presence of frustration~\cite{Sbierski2024}, such as in long-range antiferromagnets.
Matrix product state methods \cite{Crosswhite2008,Zaletel2015} are exponentially expensive with system width~\cite{StoudenmireWhite2012}.
Traditional tensor network states (TNS) approaches~\cite{xiang2023density} rely on local-gate updates~\cite{jiangAccurateDeterminationTensor2008b}, which are inefficient for long-range couplings.

Variational Monte Carlo (VMC) with TNS~\cite{Liu2021,Vieijra2021,WuDai2026,chen2026variationalmontecarlovmc} or neural quantum states (NQS)~\cite{CarleoTroyer2017,Sharir2020} avoids these restrictions: it has no sign problem and treats ground states~\cite{Sorella2001} and real-time dynamics~\cite{Carleo2017,SchmittHeyl2020,Comparin2022,Wu2026} on the same footing.
Couplings diagonal in the sampling basis, even when long-ranged, require a cheap classical sum and can be efficiently handled with conventional VMC, as represented by Refs.~\cite{heyl_prl, jannes_np, rizzi_arxiv}.
Physical systems, however, generally also host \emph{off-diagonal} long-range couplings, such as the dipolar exchange $\sigma_i^x\sigma_j^x + \sigma_i^y\sigma_j^y$, which connect each configuration to $O(N^2)$ others on $N$ sites.
Computing each coupling requires an amplitude evaluation that costs $O(N)$ for both TNS and NQS, so the energy evaluation costs $O(N^3)$. 
This is a prohibitive cost with $N$ on the order of hundreds, when benchmarking current quantum simulators or extrapolating many-body numerics to the thermodynamic limit, both of which we pursue in this paper.

Here we remove the computational obstacle for TNS by reducing the complexity from $O(N^3)$ to $O(N)$.
The key observation is that the VMC update equation is linear in the Hamiltonian, which allows us to:
(1) Decompose the Hamiltonian into a small set of components, each admitting a basis representation in which the couplings are either diagonal or local;
(2) Solve the VMC equation using the derivative obtained by summing all of the sub-derivatives that were computed separately in each basis with the requisite basis transformations.
This is a structural advantage of tensor networks, where local basis rotations are exact and cheap.
For a generic NQS, by contrast, an amplitude in a rotated basis is a sum over exponentially many computational-basis terms.
Any two-body spin-$1/2$ Hamiltonian generally requires at most three bases, and the leading cost becomes $O(N)$ TNS contractions per sample.

To demonstrate our method, we first study the ground state of the QMC-signful antiferromagnetic (AFM) $1/r^\alpha$ XY model for $\alpha=1$, $2$, $3$, and $6$.
With adaptive-basis VMC, we address the fundamental question of ordering: the N\'eel order survives the frustration for $\alpha\ge2$ but is not observed at $\alpha=1$ up to $20\times20$.
We then turn to real-time dynamics and simulate the adiabatic preparation of the dipolar antiferromagnet on the $10\times10$ Rydberg array of Ref.~\cite{Chen2023}.
There, the experimentally prepared state showed much weaker AFM correlations than the target ground state, which cast doubt on whether the adiabatic protocol can prepare this ground state in principle.
We show that the ideal protocol does prepare it, which resolves this question and places the origin of the discrepancy in the experimental realization rather than in the protocol.

\textit{Adaptive-basis VMC.}
In VMC, the variational parameters $\vec\theta$ of $\ket{\psi_{\vec\theta}}$ are updated by
\begin{equation}
    S\dot{\boldsymbol{\theta}} = -\gamma \mathbf{F},
    \label{eq:tdvp}
\end{equation}
with $\gamma = 1$ for imaginary-time evolution via stochastic reconfiguration~\cite{Sorella2001} and $\gamma = i$ for real-time evolution via time-dependent VMC (tVMC, based on the time-dependent variational principle~\cite{Haegeman2011, Carleo2012}).
The quantum geometric tensor (QGT)~\cite{ProvostVallee1980} $S$ and the force $\vec F$ are estimated as averages over $p(\vec s)\propto|\psi_{\vec\theta}(\vec s)|^2$:
\begin{align}
  S_{\mu\nu} &= \braket{O_\mu^* O_\nu} - \braket{O^*_\mu} \braket{O_\nu},\\
  F_{\mu} &= \braket{O_\mu^* H_\loc} - \braket{O^*_\mu} \braket{H_\loc},
\end{align}
with the log-derivatives $O_{\mu}(\vec s)=\partial_{\theta_\mu}\log\psi_{\vec\theta}(\vec s)$ and local energy $H_\text{loc}(\vec s) = \braket{\vec s|H|\psi}/\braket{\vec s|\psi}$. 

Two properties of Eq.~\eqref{eq:tdvp} drive the method.
First, $\vec F$ is linear in $H$, while $S$ does not depend on $H$ at all.
Second, $S$ and $\vec F$ are basis-independent quantities: they can be estimated in any product basis.
We therefore write 
\begin{equation}
    H=\sum_b H_b,
\end{equation} 
with $b$ labeling a product basis $\{\ket{\vec s_b}\}$ in which the long-range couplings in $H_b$ are diagonal, estimate $S$ and $\vec F_b$ from samples $\vec s_b\sim|\braket{\vec s_b|\psi_{\vec\theta}}|^2$ in the rotated parameters $U_b\vec\theta$, solve $S\dot{\vec\theta}_b=-\gamma\vec F_b$ for each $b$, and rotate back and add the results,
\begin{equation}
  \dot{\vec\theta}=\sum_b U_b^\dagger\,\dot{\vec\theta}_b .
  \label{eq:sum}
\end{equation}
Here $U_b$ is the single-site rotation from the reference basis to basis $b$, which is cheap for a TNS.

\begin{figure}[t]
\centering
\includegraphics[width=\linewidth]{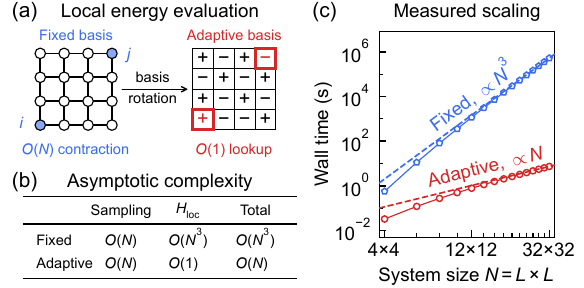}
\caption{Cost of off-diagonal long-range couplings.
(a)~In a fixed basis, an off-diagonal term $\sigma_i^a\sigma_j^a$ flips both spins $i$ and $j$, requiring $O(N)$ contractions for the amplitude of the new configuration; after rotation to the adaptive basis, it becomes diagonal and an $O(1)$ configuration lookup.
(b)~Asymptotic cost per sample in TNS contractions. 
(c)~Wall time per batch (8192 samples) in the fixed and adaptive bases for $H_{\alpha=3}$ [Eq.~\eqref{eq:Halpha}] at $D=4$, measured on a single GPU (141~GB memory, 34~FP64 TFLOPS). Dashed lines guide the eye.}
\label{fig:cost}
\end{figure}

For a general two-body spin Hamiltonian
\begin{equation}
  H=\sum_{i<j}\sum_{a,a'}J_{ij}^{aa'}\sigma_i^a\sigma_j^{a'}+\sum_i\sum_a h_i^a\sigma_i^a,
  \label{eq:general}
\end{equation}
with $a,a'\in\{x,y,z\}$, we assign each term to the basis of its first Pauli operator.
At most three bases are therefore needed, for any couplings $J^{aa'}_{ij}$.
The same-axis terms $J^{aa}_{ij}\sigma^a_i\sigma^a_j$ are then diagonal in their assigned basis, reducing to a classical pair sum $\sum_{i<j}J^{aa}_{ij}s_is_j$ with negligible cost.
Each mixed term $J^{aa'}_{ij}\sigma^a_i\sigma^{a'}_j$ reduces to an effective single-site operator  $J^{aa'}_{ij}s_i\sigma_j^{a'}$ on site $j$. 
Combined with the fields $h_i^a$, the resulting single-site terms require only $O(N)$ local contractions per sample re-using the TNS boundary environments \cite{Liu2021, WuDai2026}.
Evaluation in the adaptive basis thus reduces the cost from $O(N^3)$ to $O(N)$, on the same order as sampling itself [Fig.~\ref{fig:cost}].

\textit{Trade-offs.}
The adaptive-basis approach gives up two features of fixed-basis sampling.
First, the zero-variance principle~\cite{AssarafCaffarel1999} no longer holds, so ground-state optimization needs more samples.
Second, charge conservation can no longer be imposed by restricting the sampled sector.
Both are worthwhile at large $N$ given the scaling reduction from $O(N^3)$ to $O(N)$.

\begin{figure}[t]
\centering
\includegraphics[width=\linewidth]{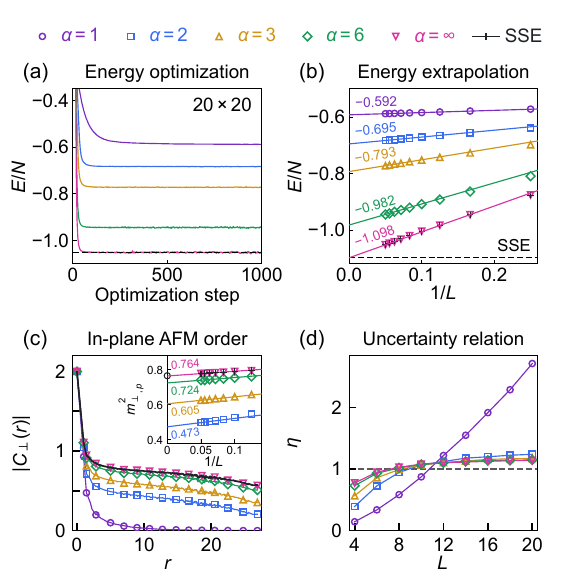}
\caption{
Ground states of $H_\alpha$.
(a)~Energy optimization on $20\times20$; the dashed line is the SSE value at $\alpha=\infty$.
(b)~$E/N$ versus $1/L$, with linear fits to $L=14$--$20$; the dashed line is the extrapolated SSE value at $\alpha=\infty$~\cite{Sandvik1999}.
(c)~In-plane correlation $|C_\perp(r)|$ on $20\times20$.
Inset: bulk-averaged staggered magnetization $m_{\perp,p}^2$ [Eq.~\eqref{eq:mpin}] with boundary pinning $h_p=0.25$, extrapolated linearly in $1/L$; the open circle is
the SSE value of Ref.~\cite{Sandvik1999}.
(d)~Uncertainty ratio $\eta$ [Eq.~\eqref{eq:robertson}]; the dashed line marks the bound $\eta=1$.
Extrapolated values in (b) and (c) are rounded to three decimal places.
}
\label{fig:alpha}
\end{figure}

\textit{AFM XY ground states.}
We begin by investigating the ground-state ordering of 
\begin{equation}
  H_\alpha=\frac{1}{2}\sum_{i<j}\frac{1}{r_{ij}^{\alpha}}\left(\sigma_i^x\sigma_j^x+\sigma_i^y\sigma_j^y\right),
  \label{eq:Halpha}
\end{equation}
for representative exponents $\alpha=1, 2, 3, 6$, and $\infty$. 
Here, $\alpha=3$ corresponds to the dipolar regime realized in Rydberg atom arrays~\cite{Chen2023}; $\alpha=6$ arises from second-order van der Waals exchange between Rydberg states~\cite{Signoles2021}; and $\alpha\lesssim3$ is realizable in trapped ions~\cite{Britton2012,Islam2013,Monroe2021}. 
While the nearest-neighbor limit $\alpha=\infty$ is sign-free and N\'eel ordered~\cite{Kennedy1988,Kubo1988,Sandvik1999}, the fate of the N\'eel order upon long-range couplings remains unclear. 
Probing this regime in 2D has hitherto been hindered by a frustration-induced QMC sign problem and an $O(N^3)$ VMC cost.
Our adaptive-basis approach overcomes these bottlenecks, reaching system sizes up to $20\times20$ to resolve the impact of the long-range frustration.

Specifically, we sample the $\sigma_i^x\sigma_j^x$ and $\sigma_i^y\sigma_j^y$ interactions in the $x$ and $y$ bases, respectively, using $81920$ samples per basis and optimization step while imposing $C_{2v}$ lattice symmetry (End Matter).
With this setup, the optimization converges within a few hundred steps for all $\alpha$ [Fig.~\ref{fig:alpha}(a)].
In the sign-free limit $\alpha=\infty$, our TNS-VMC energies agree well with stochastic series expansion (SSE) QMC~\cite{SandvikKurkijarvi1991,SyljuasenSandvik2002} across all finite lattices and in the extrapolated thermodynamic limit~\cite{Sandvik1999} [Fig.~\ref{fig:alpha}(b) and Table~\ref{tab:extrap}].

To characterize magnetic order across $\alpha$, we measure the two-point correlation function $C_a(i,j)=\braket{\sigma_i^a\sigma_j^a}$ ($a=x,y$) and its distance average $C_a(r)$ over all pairs at $r=\vert\vec r_i-\vec r_j\vert$ [$\vec r_i=(x_i,y_i)$], along with the $U(1)$-invariant combination $C_\perp=C_x+C_y$.
Our results reveal two distinct regimes [Fig.~\ref{fig:alpha}(c)].
For $\alpha\ge2$, $|C_\perp(r)|$ develops a clear plateau signifying long-range N\'eel order, before turning down at large $r$ due to open boundaries, mirroring the $\alpha=\infty$ benchmark.
In contrast, at $\alpha=1$, $|C_\perp(r)|$ decays toward zero, indicating the absence of N\'eel order.

To accurately determine the spontaneous in-plane AFM magnetization in the ordered regime ($\alpha\ge2$), we apply a staggered pinning field $h_p=0.25$ on the boundary~\cite{WhiteChernyshev2007,AssaadHerbut2013}, measure the in-plane staggered magnetization averaged over the set $C$ of the central $2\times2$ sites,
\begin{equation}
  m_{\perp,p}^2=\frac14\sum_{i\in C}\left(\braket{\sigma_i^x}^2+\braket{\sigma_i^y}^2\right),
  \label{eq:mpin}
\end{equation}
and extrapolate it linearly in $1/L$ [Fig.~\ref{fig:alpha}(c), inset, and Table~\ref{tab:extrap}]; see Supplemental Material (SM)~\cite{sm}.
At $\alpha=\infty$, the extrapolated $m_{\perp,p}^2$ agrees with the SSE benchmark~\cite{Sandvik1999}.
It decreases with decreasing $\alpha$, because the frustrating long-range tail suppresses the N\'eel order, though the order survives down to $\alpha=2$.

\begin{table}[t]
\caption{Extrapolated ground-state energy $E/N$ and staggered magnetization $m_{\perp,p}^2$ from linear fits in $1/L$ ($L=14$--$20$), with statistical
  uncertainties~\cite{sm}.}
\label{tab:extrap}
\renewcommand{\arraystretch}{1.2}
\begin{ruledtabular}
\begin{tabular}{@{\hspace{2.5em}}ccc@{\hspace{2.5em}}}
$\alpha$ & $E/N$ & $m_{\perp,p}^2$\\
\hline
1 & $-0.59161(6)$ & --\\
2 & $-0.6947(6)$ & $0.473(3)$\\
3 & $-0.793(2)$ & $0.605(3)$\\
6 & $-0.982(3)$ & $0.724(4)$\\
$\infty$ & $-1.098(4)$ & $0.764(4)$\\
\hline
$\infty$~(SSE~\cite{Sandvik1999}) & $-1.097648(4)$ & $0.765(5)$\\
\end{tabular}
\end{ruledtabular}
\end{table}

The two regimes are further distinguished by their quantum fluctuations [Fig.~\ref{fig:alpha}(d)].
The uniform in-plane moment $M_a=N^{-1/2}\sum_i\sigma_i^a$ ($a=x,y$) and the staggered out-of-plane moment $Z=N^{-1/2}\sum_i\epsilon_i\sigma_i^z$ [$\epsilon_i=(-1)^{x_i+y_i}$] do not commute, with $[M_a,Z]$ proportional to the staggered in-plane magnetization.
They therefore obey a Robertson uncertainty relation~\cite{Robertson1929,sm}, whose $U(1)$-invariant form reads
\begin{equation}
    \eta\equiv\frac{\braket{M_\perp^2}\braket{Z^2}}{m_\perp^2}\ge 1\quad (L\to\infty),
  \label{eq:robertson}
\end{equation}
where $M_\perp^2=M_x^2+M_y^2$ and $m_\perp^2=m_x^2+m_y^2$, with $m_a^2=N^{-2}\sum_{i,j}\epsilon_i\epsilon_j C_a(i,j)$ denoting the finite-size squared staggered magnetization.
For $\alpha=1$, $\eta$ grows steadily with $L$ as $m_\perp^2\to 0$.
In contrast, for $\alpha\ge2$, $\eta$ saturates slightly above unity, the exact lower bound ($\eta=1$) of a classical N\'eel state.
Although the frustrating long-range tail suppresses $m_\perp^2$ as $\alpha$ decreases, $\eta$ remains nearly invariant for $\alpha>2$ (and still close to unity at $\alpha=2$).
This indicates that long-range couplings primarily dress the quantum fluctuations without disrupting the nearly minimum-uncertainty character of the N\'eel state, until the magnetic order vanishes at $\alpha=1$ as numerically observed on the $20\times 20$ lattice.

In general, frustration strengthens as $\alpha$ decreases, which may stabilize a spin liquid below a critical $\alpha_c$. 
For $\alpha=1$, due to its exceedingly slow long-range tail, extrapolating the vanishing magnetic order from $20\times20$ to the thermodynamic limit would require substantially larger system sizes, which we do not pursue here.
We leave the question whether the ground state at $\alpha=1$ is a spin liquid to future work.

\textit{Rydberg dynamics.}
Having studied the ground-state properties of $H_\alpha$, we turn to real-time dynamics and simulate the Rydberg-atom experiment of Ref.~\cite{Chen2023} on a $10\times10$ square array partitioned into sublattices $A$ and $B$, which was also designed to study the ground-state ordering of the dipolar XY magnet.
The system evolves under
\begin{equation}
  H(t) = -H_{\alpha=3}+\delta(t)\sum_{i\in B}\frac{1+\sigma_i^z}{2},
  \label{eq:Ht}
\end{equation}
with $\delta(t)=\delta_0 e^{-t/1.45}$.
To facilitate comparison with the experiment, we measure energy in units of $J/h=0.77$~MHz and time in units of $\hbar/J=0.207\,\mu$s~\cite{sm}.

Starting from the N\'eel state $\prod_{i\in A}\ket{\uparrow}_i\prod_{i\in B}\ket{\downarrow}_i$, the staggered $\sigma^z$ field is ramped down.
For $\delta_0<0$, the ramp follows the highest-energy branch of $H(t)$. 
For the ferromagnetic protocol ($\delta_0>0$), the experimentally prepared state agreed qualitatively with the ground state of $-H_{\alpha=3}$ obtained with the density-matrix renormalization group (DMRG)~\cite{White1992, Chen2023}.
For the AFM protocol, however, the prepared state showed \emph{suppressed} correlations, whereas both DMRG~\cite{Chen2023,Bintz2026} and our ground-state results above predict \emph{strong} ones.
This discrepancy raises a natural question: can the coherent ramp actually build these ground-state correlations within the available time, or is it obstructed by nonadiabaticity of the protocol itself? 
Resolving this question demands controlled classical real-time simulations at the 2D experimental scale, which have so far been limited to short-range models~\cite{Lubasch2011,Kaneko2022,Liao2023,Tindall2024,Begusic2024,Patra2024,MauronCarleo2025,Tindall2026,Wiersema2026,Zhang2025,Haghshenas2026} and, to our knowledge, remain out of reach for systems with off-diagonal long-range couplings.

\begin{figure}[t]
\centering
\includegraphics[width=\linewidth]{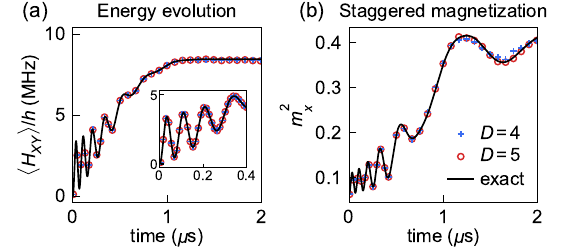}
\caption{Real-time benchmark on a $4\times4$ array against exact evolution, for the protocol of Eq.~\eqref{eq:Ht}, with bond dimensions $D=4$ and $5$ and $81920$ samples.
(a) Energy $\braket{H_\text{XY}}$, with $H_\text{XY}=-H_{\alpha=3}$.
(b) Squared staggered magnetization $m_x^2$.}
\label{fig:4x4}
\end{figure}

We address this question using adaptive-basis TNS-tVMC.
The added field term $\sigma_i^z$ [Eq.~\eqref{eq:Ht}] requires no extra sampling basis: we retain the $x$ and $y$ bases for the two-body couplings as in the ground-state case, while evaluating all single-site $\sigma_i^z$ terms in the $y$ basis, maintaining $O(N)$ scaling via TNS environment reuse \cite{Liu2021,WuDai2026}.
Since $H(t)$ and the initial state are $C_{2v}$-invariant, we enforce this symmetry on the TNS (End Matter).
We do not include device-level imperfections such as state-preparation and readout errors, decoherence, or residual non-XY couplings.
We simulate the ideal protocol starting from an exact N\'eel state with initial detuning $|\delta_0|=19.48$ ($2\pi\times15\,\mathrm{MHz}$), as realized in the $6\times7$ experiment of Ref.~\cite{Chen2023}.
For the $10\times10$ array, this $|\delta_0|$ exceeds the laser-power-limited experimental value $|\delta_0|=11.69$.
Since the N\'eel state is the exact target eigenstate of $H(0)$ only as $|\delta_0|\to\infty$, a larger $|\delta_0|$ reduces the initial-state mismatch without slowing the subsequent ramp.

Evolving from the N\'eel initial state requires additional algorithmic care. When this product state is represented by a $D>1$ TNS, its tangent space collapses to the $D=1$ subspace, rendering the QGT singular.
We therefore take the first time step with projected tVMC (p-tVMC)~\cite{Sinibaldi2023,Gravina2025}, whose target state can here be evaluated exactly and sampled directly, and continue with tVMC~\cite{Wu2026} thereafter; see End Matter.
Figure~\ref{fig:4x4} benchmarks the adaptive-basis TNS-tVMC workflow on a $4\times4$ array, where it reproduces the exact energy and staggered-magnetization trajectories.

\begin{figure}[t]
\centering
\includegraphics[width=\linewidth]{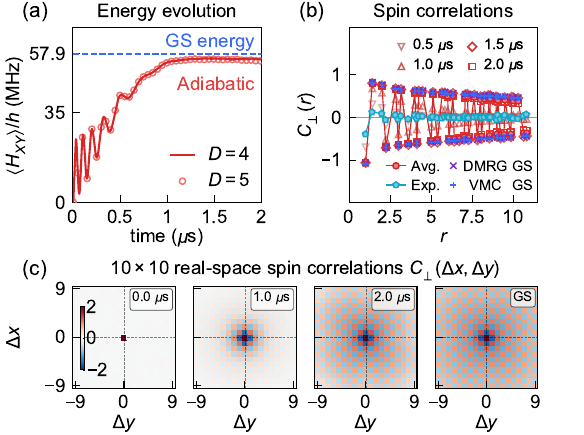}
\caption{Adiabatic preparation of the dipolar antiferromagnet on the $10\times10$ array.
(a) Energy $\braket{H_\text{XY}}$ for $D=4$ and $5$, approaching $-E_0$ (dashed line), where $E_0$ is the ground-state (GS) energy of $H_{\alpha=3}$ obtained by adaptive-basis VMC.
(b) $C_\perp(r)$ at $t=0.5$, $1.0$, $1.5$, and $2.0~\mu$s, the late-time average over $t=1.2$--$2.0~\mu$s (Avg.), and the VMC GS, compared with the experimental data (Exp.) and DMRG GS of Ref.~\cite{Chen2023}, for which $C_\perp\simeq2C_x$ by $U(1)$ symmetry.
(c) Real-space correlations $C_\perp(\Delta x,\Delta y)$, averaged over pairs with displacement $(\Delta x,\Delta y)$, at $t=0$, $1.0$, and $2.0~\mu$s and in the VMC GS.
}
\label{fig:10x10}
\end{figure}

On the $10\times10$ array [Fig.~\ref{fig:10x10}(a)], the energy exhibits rapid early-time oscillations before settling for $t\gtrsim1.2\,\mu$s onto a plateau close to the target ground-state value obtained with TNS-VMC.
The in-plane correlation $C_\perp$ builds up during the ramp and, at its end, closely follows the ground-state profile at all distances [Fig.~\ref{fig:10x10}(b)]; the full real-space correlation maps of the dynamical and target states show essentially the same extended AFM pattern [Fig.~\ref{fig:10x10}(c)].
These results are converged in bond dimension and sample size (see SM~\cite{sm}).

This answers the question posed by the experiment: the ideal protocol does prepare the AFM ground state of $H_{\alpha=3}$ on the $10\times10$ array, including its spatially extended correlations, within the experimental ramp time.
The long-range interactions and finite ramp time thus pose no intrinsic obstruction to adiabatic preparation.
The suppressed correlations observed experimentally therefore stem from departures from the ideal protocol.

\textit{Outlook.}
Adaptive-basis sampling makes general two-dimensional long-range quantum spin systems accessible to tensor-network VMC, for both ground states and real-time dynamics.
As a numerical method, it applies to a variety of long-ranged many-body systems due to the generality of Eq.~\eqref{eq:general} at large system size. 
As a classical simulator of quantum simulators, it can benchmark, interpret, and, more importantly, guide the real-time protocol for current Rydberg and ion experiments. 
The limits of the method are those of the tensor network itself: the method is most powerful for ground states, adiabatic and low-energy dynamics, and local quenches, while global quenches generate entanglement rapidly and limit the reachable times, which is also a natural relay point between classical and quantum simulation. 
Long-range fermionic hopping lies outside the present scheme and is an exciting future direction. 

\textit{Acknowledgments.}
Part of the algorithm was tested using TeNPy~\cite{HauschildPollmann2018,TeNPy2024}, and the production code was implemented in JAX~\cite{JAX2018}.
Y.W. is grateful for discussions with his colleagues Cheng Chen and Shi-Xin Zhang.
Y.W. thanks Jannes Nys for collaboration on a closely related project, and in particular for the p-tVMC codes shared in the collaboration.
Y.W. thanks Tao Chen for sharing the code for column direct sampling of PEPS during the collaboration of Ref.~\cite{chen2026variationalmontecarlovmc}.
The data supporting this work are openly available in Ref.~\cite{Data2026}; embargo periods may apply.
Y.W. acknowledges support from the Chinese Academy of Sciences (CAS) under Grant No. YSBR-150 and a start-up grant from IOP-CAS.
This work was partly supported by the National Natural Science Foundation of China (Grant No. 12488201) and the National Key Research and Development Program of China (2021ZD0301800).
The simulations were supported by the IOP-CAS computing facilities.

\section*{End Matter}

\textit{Lattice symmetry.}
The Hamiltonians and the N\'eel initial state are invariant under reflections about both diagonals of the square lattice, which generate the point group $C_{2v}$.
We therefore impose it on the ansatz by tying tensors related by the reflections.
This reduces the number of independent parameters by roughly a factor of four, lowering the GPU memory and the variance of $\dot{\vec\theta}$ at a fixed number of samples.

\textit{Evolving from product states.}
At an exact product state, the tangent space of a TNS with $D>1$ reduces to that of product states: the QGT $S$ acquires null directions, leading to a pathology of the TDVP.
We therefore use p-tVMC~\cite{Sinibaldi2023,Gravina2025} for the first time step only, starting from the N\'eel state
$\ket{\phi}=\prod_{i\in A}\ket{\uparrow}_i\prod_{i\in B}\ket{\downarrow}_i$.
Instead of Eq.~\eqref{eq:tdvp}, we determine $\vec\theta(\delta t)$ by maximizing the fidelity
\begin{equation}
\vec\theta(\delta t)=\underset{\vec\theta}{\arg\max}\;
\frac{\abs{\braket{\psi_{\vec\theta}|\phi'}}^2}
{\braket{\psi_{\vec\theta}|\psi_{\vec\theta}}\braket{\phi'|\phi'}}
\label{eq:em_ptvmc}
\end{equation}
with the second-order short-time target
\begin{equation}
\begin{aligned}
  \ket{\phi'}&=V_2(0)\ket{\phi},\\
  V_2(t)&=1-iH(t)\delta t-\frac{\delta t^2}{2}\left[H(t)^2+i\dot H(t)\right],
\end{aligned}
\label{eq:em_ptvmc_target}
\end{equation}
where $H(t)=H_{XY}+H_Z(t)$, with $H_{XY}=-H_{\alpha=3}$ and $H_Z(t)=\delta(t)\sum_{i\in B}(1+\sigma_i^z)/2$.
Unlike Eq.~\eqref{eq:tdvp}, Eq.~\eqref{eq:em_ptvmc} is a nonlinear optimization that is not restricted to the tangent space at $\ket{\phi}$: $\ket{\phi'}$ contains entangled components, which a TNS with $D>1$ can represent.
After this single step, $S$ is regular, and we revert to tVMC~\cite{Wu2026}.

Although p-tVMC is generally expensive, its target is cheap here.
The fidelity is estimated with samples from
\begin{equation}
  p_2(\mathbf z)=
  \frac{|\Phi_2(\mathbf z)|^2}
  {\sum_{\mathbf z}|\Phi_2(\mathbf z)|^2},
  \qquad
  \Phi_2(\mathbf z)\equiv\braket{\mathbf z|\phi'},
  \label{eq:em_ptvmc_distribution}
\end{equation}
which are drawn directly in the $z$ basis, without any tensor-network contraction.
Let $\mathbf z_0$ denote the N\'eel configuration, let $\mathbf z^{ij}$ be obtained from $\mathbf z$ by flipping spins $i$ and $j$, and write $J_{ij}=r_{ij}^{-3}$.
The Hamiltonian acts sparsely as
\begin{align}
 H_{XY}\ket{\mathbf z}
 &= -\sum_{i<j\, :\, z_i=-z_j}J_{ij}\ket{\mathbf z^{ij}},
 \label{eq:em_xy_sparse_action}\\
 H_Z(t)\ket{\mathbf z}
 &=\delta(t)\sum_{i\in B}\frac{1+z_i}{2}\,\ket{\mathbf z}.
 \label{eq:em_z_sparse_action}
\end{align}
Since $H_Z(0)\ket{\phi}=\dot H_Z(0)\ket{\phi}=0$, the target amplitude is
\begin{equation}
 \Phi_2(\mathbf z)=
 \delta_{\mathbf z,\mathbf z_0}
 -i\delta t\,h_1(\mathbf z)
 -\frac{\delta t^2}{2}\,h_2(\mathbf z),
 \label{eq:em_ptvmc_amplitude}
\end{equation}
with
\begin{equation}
\begin{aligned}
 h_1(\mathbf z)&=\braket{\mathbf z|H_{XY}|\mathbf z_0},\\
 h_2(\mathbf z)&=\braket{\mathbf z|\big(H_{XY}^2+H_Z(0)H_{XY}\big)|\mathbf z_0}.
\end{aligned}
\end{equation}
Starting from $\mathbf z_0$, one application of Eq.~\eqref{eq:em_xy_sparse_action} gives $h_1$, and applying Eq.~\eqref{eq:em_xy_sparse_action} and the diagonal factor of Eq.~\eqref{eq:em_z_sparse_action} to $h_1$ gives $h_2$, so $\Phi_2$ is evaluated exactly and is supported only on configurations at most two XY exchanges away from $\mathbf z_0$.
We sample Eq.~\eqref{eq:em_ptvmc_distribution} with a Metropolis chain whose proposals flip a uniformly selected pair of spins and are accepted with probability $\min\{1,|\Phi_2(\mathbf z')|^2/|\Phi_2(\mathbf z)|^2\}$; proposals outside the support of $\Phi_2$ are rejected automatically.
These samples enter the p-tVMC optimization of Eq.~\eqref{eq:em_ptvmc} following Refs.~\cite{Sinibaldi2023,Gravina2025}.

\bibliography{ref}

\begin{thebibliography}{71}%
\makeatletter
\providecommand \@ifxundefined [1]{%
 \@ifx{#1\undefined}
}%
\providecommand \@ifnum [1]{%
 \ifnum #1\expandafter \@firstoftwo
 \else \expandafter \@secondoftwo
 \fi
}%
\providecommand \@ifx [1]{%
 \ifx #1\expandafter \@firstoftwo
 \else \expandafter \@secondoftwo
 \fi
}%
\providecommand \natexlab [1]{#1}%
\providecommand \enquote  [1]{``#1''}%
\providecommand \bibnamefont  [1]{#1}%
\providecommand \bibfnamefont [1]{#1}%
\providecommand \citenamefont [1]{#1}%
\providecommand \href@noop [0]{\@secondoftwo}%
\providecommand \href [0]{\begingroup \@sanitize@url \@href}%
\providecommand \@href[1]{\@@startlink{#1}\@@href}%
\providecommand \@@href[1]{\endgroup#1\@@endlink}%
\providecommand \@sanitize@url [0]{\catcode `\\12\catcode `\$12\catcode
  `\&12\catcode `\#12\catcode `\^12\catcode `\_12\catcode `\%12\relax}%
\providecommand \@@startlink[1]{}%
\providecommand \@@endlink[0]{}%
\providecommand \url  [0]{\begingroup\@sanitize@url \@url }%
\providecommand \@url [1]{\endgroup\@href {#1}{\urlprefix }}%
\providecommand \urlprefix  [0]{URL }%
\providecommand \Eprint [0]{\href }%
\providecommand \doibase [0]{https://doi.org/}%
\providecommand \selectlanguage [0]{\@gobble}%
\providecommand \bibinfo  [0]{\@secondoftwo}%
\providecommand \bibfield  [0]{\@secondoftwo}%
\providecommand \translation [1]{[#1]}%
\providecommand \BibitemOpen [0]{}%
\providecommand \bibitemStop [0]{}%
\providecommand \bibitemNoStop [0]{.\EOS\space}%
\providecommand \EOS [0]{\spacefactor3000\relax}%
\providecommand \BibitemShut  [1]{\csname bibitem#1\endcsname}%
\let\auto@bib@innerbib\@empty
\bibitem [{\citenamefont {de~Paz}\ \emph {et~al.}(2013)\citenamefont {de~Paz},
  \citenamefont {Sharma}, \citenamefont {Chotia}, \citenamefont {Mar{\'e}chal},
  \citenamefont {Huckans}, \citenamefont {Pedri}, \citenamefont {Santos},
  \citenamefont {Gorceix}, \citenamefont {Vernac},\ and\ \citenamefont
  {Laburthe-Tolra}}]{dePaz2013}%
  \BibitemOpen
  \bibfield  {author} {\bibinfo {author} {\bibfnamefont {A.}~\bibnamefont
  {de~Paz}}, \bibinfo {author} {\bibfnamefont {A.}~\bibnamefont {Sharma}},
  \bibinfo {author} {\bibfnamefont {A.}~\bibnamefont {Chotia}}, \bibinfo
  {author} {\bibfnamefont {E.}~\bibnamefont {Mar{\'e}chal}}, \bibinfo {author}
  {\bibfnamefont {J.~H.}\ \bibnamefont {Huckans}}, \bibinfo {author}
  {\bibfnamefont {P.}~\bibnamefont {Pedri}}, \bibinfo {author} {\bibfnamefont
  {L.}~\bibnamefont {Santos}}, \bibinfo {author} {\bibfnamefont
  {O.}~\bibnamefont {Gorceix}}, \bibinfo {author} {\bibfnamefont
  {L.}~\bibnamefont {Vernac}},\ and\ \bibinfo {author} {\bibfnamefont
  {B.}~\bibnamefont {Laburthe-Tolra}},\ }\bibfield  {title} {\bibinfo {title}
  {Nonequilibrium quantum magnetism in a dipolar lattice gas},\ }\href
  {https://doi.org/10.1103/physrevlett.111.185305} {\bibfield  {journal}
  {\bibinfo  {journal} {Phys. Rev. Lett.}\ }\textbf {\bibinfo {volume} {111}},\
  \bibinfo {pages} {185305} (\bibinfo {year} {2013})}\BibitemShut {NoStop}%
\bibitem [{\citenamefont {Gorshkov}\ \emph {et~al.}(2011)\citenamefont
  {Gorshkov}, \citenamefont {Manmana}, \citenamefont {Chen}, \citenamefont
  {Ye}, \citenamefont {Demler}, \citenamefont {Lukin},\ and\ \citenamefont
  {Rey}}]{Gorshkov2011}%
  \BibitemOpen
  \bibfield  {author} {\bibinfo {author} {\bibfnamefont {A.~V.}\ \bibnamefont
  {Gorshkov}}, \bibinfo {author} {\bibfnamefont {S.~R.}\ \bibnamefont
  {Manmana}}, \bibinfo {author} {\bibfnamefont {G.}~\bibnamefont {Chen}},
  \bibinfo {author} {\bibfnamefont {J.}~\bibnamefont {Ye}}, \bibinfo {author}
  {\bibfnamefont {E.}~\bibnamefont {Demler}}, \bibinfo {author} {\bibfnamefont
  {M.~D.}\ \bibnamefont {Lukin}},\ and\ \bibinfo {author} {\bibfnamefont
  {A.~M.}\ \bibnamefont {Rey}},\ }\bibfield  {title} {\bibinfo {title} {Tunable
  superfluidity and quantum magnetism with ultracold polar molecules},\ }\href
  {https://doi.org/10.1103/physrevlett.107.115301} {\bibfield  {journal}
  {\bibinfo  {journal} {Phys. Rev. Lett.}\ }\textbf {\bibinfo {volume} {107}},\
  \bibinfo {pages} {115301} (\bibinfo {year} {2011})}\BibitemShut {NoStop}%
\bibitem [{\citenamefont {Yan}\ \emph {et~al.}(2013)\citenamefont {Yan},
  \citenamefont {Moses}, \citenamefont {Gadway}, \citenamefont {Covey},
  \citenamefont {Hazzard}, \citenamefont {Rey}, \citenamefont {Jin},\ and\
  \citenamefont {Ye}}]{Yan2013}%
  \BibitemOpen
  \bibfield  {author} {\bibinfo {author} {\bibfnamefont {B.}~\bibnamefont
  {Yan}}, \bibinfo {author} {\bibfnamefont {S.~A.}\ \bibnamefont {Moses}},
  \bibinfo {author} {\bibfnamefont {B.}~\bibnamefont {Gadway}}, \bibinfo
  {author} {\bibfnamefont {J.~P.}\ \bibnamefont {Covey}}, \bibinfo {author}
  {\bibfnamefont {K.~R.~A.}\ \bibnamefont {Hazzard}}, \bibinfo {author}
  {\bibfnamefont {A.~M.}\ \bibnamefont {Rey}}, \bibinfo {author} {\bibfnamefont
  {D.~S.}\ \bibnamefont {Jin}},\ and\ \bibinfo {author} {\bibfnamefont
  {J.}~\bibnamefont {Ye}},\ }\bibfield  {title} {\bibinfo {title} {Observation
  of dipolar spin-exchange interactions with lattice-confined polar
  molecules},\ }\href {https://doi.org/10.1038/nature12483} {\bibfield
  {journal} {\bibinfo  {journal} {Nature}\ }\textbf {\bibinfo {volume} {501}},\
  \bibinfo {pages} {521} (\bibinfo {year} {2013})}\BibitemShut {NoStop}%
\bibitem [{\citenamefont {Christakis}\ \emph {et~al.}(2023)\citenamefont
  {Christakis}, \citenamefont {Rosenberg}, \citenamefont {Raj}, \citenamefont
  {Chi}, \citenamefont {Morningstar}, \citenamefont {Huse}, \citenamefont
  {Yan},\ and\ \citenamefont {Bakr}}]{Christakis2023}%
  \BibitemOpen
  \bibfield  {author} {\bibinfo {author} {\bibfnamefont {L.}~\bibnamefont
  {Christakis}}, \bibinfo {author} {\bibfnamefont {J.~S.}\ \bibnamefont
  {Rosenberg}}, \bibinfo {author} {\bibfnamefont {R.}~\bibnamefont {Raj}},
  \bibinfo {author} {\bibfnamefont {S.}~\bibnamefont {Chi}}, \bibinfo {author}
  {\bibfnamefont {A.}~\bibnamefont {Morningstar}}, \bibinfo {author}
  {\bibfnamefont {D.~A.}\ \bibnamefont {Huse}}, \bibinfo {author}
  {\bibfnamefont {Z.~Z.}\ \bibnamefont {Yan}},\ and\ \bibinfo {author}
  {\bibfnamefont {W.~S.}\ \bibnamefont {Bakr}},\ }\bibfield  {title} {\bibinfo
  {title} {Probing site-resolved correlations in a spin system of ultracold
  molecules},\ }\href {https://doi.org/10.1038/s41586-022-05558-4} {\bibfield
  {journal} {\bibinfo  {journal} {Nature}\ }\textbf {\bibinfo {volume} {614}},\
  \bibinfo {pages} {64} (\bibinfo {year} {2023})}\BibitemShut {NoStop}%
\bibitem [{\citenamefont {Zu}\ \emph {et~al.}(2021)\citenamefont {Zu},
  \citenamefont {Machado}, \citenamefont {Ye}, \citenamefont {Choi},
  \citenamefont {Kobrin}, \citenamefont {Mittiga}, \citenamefont {Hsieh},
  \citenamefont {Bhattacharyya}, \citenamefont {Markham}, \citenamefont
  {Twitchen}, \citenamefont {Jarmola}, \citenamefont {Budker}, \citenamefont
  {Laumann}, \citenamefont {Moore},\ and\ \citenamefont {Yao}}]{Zu2021}%
  \BibitemOpen
  \bibfield  {author} {\bibinfo {author} {\bibfnamefont {C.}~\bibnamefont
  {Zu}}, \bibinfo {author} {\bibfnamefont {F.}~\bibnamefont {Machado}},
  \bibinfo {author} {\bibfnamefont {B.}~\bibnamefont {Ye}}, \bibinfo {author}
  {\bibfnamefont {S.}~\bibnamefont {Choi}}, \bibinfo {author} {\bibfnamefont
  {B.}~\bibnamefont {Kobrin}}, \bibinfo {author} {\bibfnamefont
  {T.}~\bibnamefont {Mittiga}}, \bibinfo {author} {\bibfnamefont
  {S.}~\bibnamefont {Hsieh}}, \bibinfo {author} {\bibfnamefont
  {P.}~\bibnamefont {Bhattacharyya}}, \bibinfo {author} {\bibfnamefont
  {M.}~\bibnamefont {Markham}}, \bibinfo {author} {\bibfnamefont
  {D.}~\bibnamefont {Twitchen}}, \bibinfo {author} {\bibfnamefont
  {A.}~\bibnamefont {Jarmola}}, \bibinfo {author} {\bibfnamefont
  {D.}~\bibnamefont {Budker}}, \bibinfo {author} {\bibfnamefont {C.~R.}\
  \bibnamefont {Laumann}}, \bibinfo {author} {\bibfnamefont {J.~E.}\
  \bibnamefont {Moore}},\ and\ \bibinfo {author} {\bibfnamefont {N.~Y.}\
  \bibnamefont {Yao}},\ }\bibfield  {title} {\bibinfo {title} {Emergent
  hydrodynamics in a strongly interacting dipolar spin ensemble},\ }\href
  {https://doi.org/10.1038/s41586-021-03763-1} {\bibfield  {journal} {\bibinfo
  {journal} {Nature}\ }\textbf {\bibinfo {volume} {597}},\ \bibinfo {pages}
  {45} (\bibinfo {year} {2021})}\BibitemShut {NoStop}%
\bibitem [{\citenamefont {Defenu}\ \emph {et~al.}(2023)\citenamefont {Defenu},
  \citenamefont {Donner}, \citenamefont {Macr{\`i}}, \citenamefont {Pagano},
  \citenamefont {Ruffo},\ and\ \citenamefont {Trombettoni}}]{Defenu2023}%
  \BibitemOpen
  \bibfield  {author} {\bibinfo {author} {\bibfnamefont {N.}~\bibnamefont
  {Defenu}}, \bibinfo {author} {\bibfnamefont {T.}~\bibnamefont {Donner}},
  \bibinfo {author} {\bibfnamefont {T.}~\bibnamefont {Macr{\`i}}}, \bibinfo
  {author} {\bibfnamefont {G.}~\bibnamefont {Pagano}}, \bibinfo {author}
  {\bibfnamefont {S.}~\bibnamefont {Ruffo}},\ and\ \bibinfo {author}
  {\bibfnamefont {A.}~\bibnamefont {Trombettoni}},\ }\bibfield  {title}
  {\bibinfo {title} {Long-range interacting quantum systems},\ }\href
  {https://doi.org/10.1103/RevModPhys.95.035002} {\bibfield  {journal}
  {\bibinfo  {journal} {Rev. Mod. Phys.}\ }\textbf {\bibinfo {volume} {95}},\
  \bibinfo {pages} {035002} (\bibinfo {year} {2023})}\BibitemShut {NoStop}%
\bibitem [{\citenamefont {Saffman}\ \emph {et~al.}(2010)\citenamefont
  {Saffman}, \citenamefont {Walker},\ and\ \citenamefont
  {M{\o}lmer}}]{SaffmanWalkerMolmer2010}%
  \BibitemOpen
  \bibfield  {author} {\bibinfo {author} {\bibfnamefont {M.}~\bibnamefont
  {Saffman}}, \bibinfo {author} {\bibfnamefont {T.~G.}\ \bibnamefont
  {Walker}},\ and\ \bibinfo {author} {\bibfnamefont {K.}~\bibnamefont
  {M{\o}lmer}},\ }\bibfield  {title} {\bibinfo {title} {Quantum information
  with {R}ydberg atoms},\ }\href {https://doi.org/10.1103/RevModPhys.82.2313}
  {\bibfield  {journal} {\bibinfo  {journal} {Rev. Mod. Phys.}\ }\textbf
  {\bibinfo {volume} {82}},\ \bibinfo {pages} {2313} (\bibinfo {year}
  {2010})}\BibitemShut {NoStop}%
\bibitem [{\citenamefont {Browaeys}\ and\ \citenamefont
  {Lahaye}(2020)}]{BrowaeysLahaye2020}%
  \BibitemOpen
  \bibfield  {author} {\bibinfo {author} {\bibfnamefont {A.}~\bibnamefont
  {Browaeys}}\ and\ \bibinfo {author} {\bibfnamefont {T.}~\bibnamefont
  {Lahaye}},\ }\bibfield  {title} {\bibinfo {title} {Many-body physics with
  individually controlled {R}ydberg atoms},\ }\href
  {https://doi.org/10.1038/s41567-019-0733-z} {\bibfield  {journal} {\bibinfo
  {journal} {Nat. Phys.}\ }\textbf {\bibinfo {volume} {16}},\ \bibinfo {pages}
  {132} (\bibinfo {year} {2020})}\BibitemShut {NoStop}%
\bibitem [{\citenamefont {Porras}\ and\ \citenamefont
  {Cirac}(2004)}]{PorrasCirac2004}%
  \BibitemOpen
  \bibfield  {author} {\bibinfo {author} {\bibfnamefont {D.}~\bibnamefont
  {Porras}}\ and\ \bibinfo {author} {\bibfnamefont {J.~I.}\ \bibnamefont
  {Cirac}},\ }\bibfield  {title} {\bibinfo {title} {Effective quantum spin
  systems with trapped ions},\ }\href
  {https://doi.org/10.1103/PhysRevLett.92.207901} {\bibfield  {journal}
  {\bibinfo  {journal} {Phys. Rev. Lett.}\ }\textbf {\bibinfo {volume} {92}},\
  \bibinfo {pages} {207901} (\bibinfo {year} {2004})}\BibitemShut {NoStop}%
\bibitem [{\citenamefont {Kim}\ \emph {et~al.}(2009)\citenamefont {Kim},
  \citenamefont {Chang}, \citenamefont {Islam}, \citenamefont {Korenblit},
  \citenamefont {Duan},\ and\ \citenamefont {Monroe}}]{Kim2009}%
  \BibitemOpen
  \bibfield  {author} {\bibinfo {author} {\bibfnamefont {K.}~\bibnamefont
  {Kim}}, \bibinfo {author} {\bibfnamefont {M.-S.}\ \bibnamefont {Chang}},
  \bibinfo {author} {\bibfnamefont {R.}~\bibnamefont {Islam}}, \bibinfo
  {author} {\bibfnamefont {S.}~\bibnamefont {Korenblit}}, \bibinfo {author}
  {\bibfnamefont {L.-M.}\ \bibnamefont {Duan}},\ and\ \bibinfo {author}
  {\bibfnamefont {C.}~\bibnamefont {Monroe}},\ }\bibfield  {title} {\bibinfo
  {title} {Entanglement and tunable spin-spin couplings between trapped ions
  using multiple transverse modes},\ }\href
  {https://doi.org/10.1103/PhysRevLett.103.120502} {\bibfield  {journal}
  {\bibinfo  {journal} {Phys. Rev. Lett.}\ }\textbf {\bibinfo {volume} {103}},\
  \bibinfo {pages} {120502} (\bibinfo {year} {2009})}\BibitemShut {NoStop}%
\bibitem [{\citenamefont {Monroe}\ \emph {et~al.}(2021)\citenamefont {Monroe},
  \citenamefont {Campbell}, \citenamefont {Duan}, \citenamefont {Gong},
  \citenamefont {Gorshkov}, \citenamefont {Hess}, \citenamefont {Islam},
  \citenamefont {Kim}, \citenamefont {Linke}, \citenamefont {Pagano},
  \citenamefont {Richerme}, \citenamefont {Senko},\ and\ \citenamefont
  {Yao}}]{Monroe2021}%
  \BibitemOpen
  \bibfield  {author} {\bibinfo {author} {\bibfnamefont {C.}~\bibnamefont
  {Monroe}}, \bibinfo {author} {\bibfnamefont {W.~C.}\ \bibnamefont
  {Campbell}}, \bibinfo {author} {\bibfnamefont {L.}~\bibnamefont {Duan}},
  \bibinfo {author} {\bibfnamefont {Z.}~\bibnamefont {Gong}}, \bibinfo {author}
  {\bibfnamefont {A.~V.}\ \bibnamefont {Gorshkov}}, \bibinfo {author}
  {\bibfnamefont {P.~W.}\ \bibnamefont {Hess}}, \bibinfo {author}
  {\bibfnamefont {R.}~\bibnamefont {Islam}}, \bibinfo {author} {\bibfnamefont
  {K.}~\bibnamefont {Kim}}, \bibinfo {author} {\bibfnamefont {N.~M.}\
  \bibnamefont {Linke}}, \bibinfo {author} {\bibfnamefont {G.}~\bibnamefont
  {Pagano}}, \bibinfo {author} {\bibfnamefont {P.}~\bibnamefont {Richerme}},
  \bibinfo {author} {\bibfnamefont {C.}~\bibnamefont {Senko}},\ and\ \bibinfo
  {author} {\bibfnamefont {N.~Y.}\ \bibnamefont {Yao}},\ }\bibfield  {title}
  {\bibinfo {title} {Programmable quantum simulations of spin systems with
  trapped ions},\ }\href {https://doi.org/10.1103/RevModPhys.93.025001}
  {\bibfield  {journal} {\bibinfo  {journal} {Rev. Mod. Phys.}\ }\textbf
  {\bibinfo {volume} {93}},\ \bibinfo {pages} {025001} (\bibinfo {year}
  {2021})}\BibitemShut {NoStop}%
\bibitem [{\citenamefont {Britton}\ \emph {et~al.}(2012)\citenamefont
  {Britton}, \citenamefont {Sawyer}, \citenamefont {Keith}, \citenamefont
  {Wang}, \citenamefont {Freericks}, \citenamefont {Uys}, \citenamefont
  {Biercuk},\ and\ \citenamefont {Bollinger}}]{Britton2012}%
  \BibitemOpen
  \bibfield  {author} {\bibinfo {author} {\bibfnamefont {J.~W.}\ \bibnamefont
  {Britton}}, \bibinfo {author} {\bibfnamefont {B.~C.}\ \bibnamefont {Sawyer}},
  \bibinfo {author} {\bibfnamefont {A.~C.}\ \bibnamefont {Keith}}, \bibinfo
  {author} {\bibfnamefont {C.-C.~J.}\ \bibnamefont {Wang}}, \bibinfo {author}
  {\bibfnamefont {J.~K.}\ \bibnamefont {Freericks}}, \bibinfo {author}
  {\bibfnamefont {H.}~\bibnamefont {Uys}}, \bibinfo {author} {\bibfnamefont
  {M.~J.}\ \bibnamefont {Biercuk}},\ and\ \bibinfo {author} {\bibfnamefont
  {J.~J.}\ \bibnamefont {Bollinger}},\ }\bibfield  {title} {\bibinfo {title}
  {Engineered two-dimensional {I}sing interactions in a trapped-ion quantum
  simulator with hundreds of spins},\ }\href
  {https://doi.org/10.1038/nature10981} {\bibfield  {journal} {\bibinfo
  {journal} {Nature}\ }\textbf {\bibinfo {volume} {484}},\ \bibinfo {pages}
  {489} (\bibinfo {year} {2012})}\BibitemShut {NoStop}%
\bibitem [{\citenamefont {Bohnet}\ \emph {et~al.}(2016)\citenamefont {Bohnet},
  \citenamefont {Sawyer}, \citenamefont {Britton}, \citenamefont {Wall},
  \citenamefont {Rey}, \citenamefont {Foss-Feig},\ and\ \citenamefont
  {Bollinger}}]{Bohnet2016}%
  \BibitemOpen
  \bibfield  {author} {\bibinfo {author} {\bibfnamefont {J.~G.}\ \bibnamefont
  {Bohnet}}, \bibinfo {author} {\bibfnamefont {B.~C.}\ \bibnamefont {Sawyer}},
  \bibinfo {author} {\bibfnamefont {J.~W.}\ \bibnamefont {Britton}}, \bibinfo
  {author} {\bibfnamefont {M.~L.}\ \bibnamefont {Wall}}, \bibinfo {author}
  {\bibfnamefont {A.~M.}\ \bibnamefont {Rey}}, \bibinfo {author} {\bibfnamefont
  {M.}~\bibnamefont {Foss-Feig}},\ and\ \bibinfo {author} {\bibfnamefont
  {J.~J.}\ \bibnamefont {Bollinger}},\ }\bibfield  {title} {\bibinfo {title}
  {Quantum spin dynamics and entanglement generation with hundreds of trapped
  ions},\ }\href {https://doi.org/10.1126/science.aad9958} {\bibfield
  {journal} {\bibinfo  {journal} {Science}\ }\textbf {\bibinfo {volume}
  {352}},\ \bibinfo {pages} {1297} (\bibinfo {year} {2016})}\BibitemShut
  {NoStop}%
\bibitem [{\citenamefont {Chen}\ \emph {et~al.}(2023)\citenamefont {Chen},
  \citenamefont {Bornet}, \citenamefont {Bintz}, \citenamefont {Emperauger},
  \citenamefont {Leclerc}, \citenamefont {Liu}, \citenamefont {Scholl},
  \citenamefont {Barredo}, \citenamefont {Hauschild}, \citenamefont
  {Chatterjee}, \citenamefont {Schuler}, \citenamefont {L{\"a}uchli},
  \citenamefont {Zaletel}, \citenamefont {Lahaye}, \citenamefont {Yao},\ and\
  \citenamefont {Browaeys}}]{Chen2023}%
  \BibitemOpen
  \bibfield  {author} {\bibinfo {author} {\bibfnamefont {C.}~\bibnamefont
  {Chen}}, \bibinfo {author} {\bibfnamefont {G.}~\bibnamefont {Bornet}},
  \bibinfo {author} {\bibfnamefont {M.}~\bibnamefont {Bintz}}, \bibinfo
  {author} {\bibfnamefont {G.}~\bibnamefont {Emperauger}}, \bibinfo {author}
  {\bibfnamefont {L.}~\bibnamefont {Leclerc}}, \bibinfo {author} {\bibfnamefont
  {V.~S.}\ \bibnamefont {Liu}}, \bibinfo {author} {\bibfnamefont
  {P.}~\bibnamefont {Scholl}}, \bibinfo {author} {\bibfnamefont
  {D.}~\bibnamefont {Barredo}}, \bibinfo {author} {\bibfnamefont
  {J.}~\bibnamefont {Hauschild}}, \bibinfo {author} {\bibfnamefont
  {S.}~\bibnamefont {Chatterjee}}, \bibinfo {author} {\bibfnamefont
  {M.}~\bibnamefont {Schuler}}, \bibinfo {author} {\bibfnamefont {A.~M.}\
  \bibnamefont {L{\"a}uchli}}, \bibinfo {author} {\bibfnamefont {M.~P.}\
  \bibnamefont {Zaletel}}, \bibinfo {author} {\bibfnamefont {T.}~\bibnamefont
  {Lahaye}}, \bibinfo {author} {\bibfnamefont {N.~Y.}\ \bibnamefont {Yao}},\
  and\ \bibinfo {author} {\bibfnamefont {A.}~\bibnamefont {Browaeys}},\
  }\bibfield  {title} {\bibinfo {title} {Continuous symmetry breaking in a
  two-dimensional {R}ydberg array},\ }\href
  {https://doi.org/10.1038/s41586-023-05859-2} {\bibfield  {journal} {\bibinfo
  {journal} {Nature}\ }\textbf {\bibinfo {volume} {616}},\ \bibinfo {pages}
  {691} (\bibinfo {year} {2023})}\BibitemShut {NoStop}%
\bibitem [{\citenamefont {Bornet}\ \emph {et~al.}(2023)\citenamefont {Bornet},
  \citenamefont {Emperauger}, \citenamefont {Chen}, \citenamefont {Ye},
  \citenamefont {Block}, \citenamefont {Bintz}, \citenamefont {Boyd},
  \citenamefont {Barredo}, \citenamefont {Comparin}, \citenamefont {Mezzacapo},
  \citenamefont {Roscilde}, \citenamefont {Lahaye}, \citenamefont {Yao},\ and\
  \citenamefont {Browaeys}}]{Bornet2023}%
  \BibitemOpen
  \bibfield  {author} {\bibinfo {author} {\bibfnamefont {G.}~\bibnamefont
  {Bornet}}, \bibinfo {author} {\bibfnamefont {G.}~\bibnamefont {Emperauger}},
  \bibinfo {author} {\bibfnamefont {C.}~\bibnamefont {Chen}}, \bibinfo {author}
  {\bibfnamefont {B.}~\bibnamefont {Ye}}, \bibinfo {author} {\bibfnamefont
  {M.}~\bibnamefont {Block}}, \bibinfo {author} {\bibfnamefont
  {M.}~\bibnamefont {Bintz}}, \bibinfo {author} {\bibfnamefont {J.~A.}\
  \bibnamefont {Boyd}}, \bibinfo {author} {\bibfnamefont {D.}~\bibnamefont
  {Barredo}}, \bibinfo {author} {\bibfnamefont {T.}~\bibnamefont {Comparin}},
  \bibinfo {author} {\bibfnamefont {F.}~\bibnamefont {Mezzacapo}}, \bibinfo
  {author} {\bibfnamefont {T.}~\bibnamefont {Roscilde}}, \bibinfo {author}
  {\bibfnamefont {T.}~\bibnamefont {Lahaye}}, \bibinfo {author} {\bibfnamefont
  {N.~Y.}\ \bibnamefont {Yao}},\ and\ \bibinfo {author} {\bibfnamefont
  {A.}~\bibnamefont {Browaeys}},\ }\bibfield  {title} {\bibinfo {title}
  {Scalable spin squeezing in a dipolar {R}ydberg atom array},\ }\href
  {https://doi.org/10.1038/s41586-023-06414-9} {\bibfield  {journal} {\bibinfo
  {journal} {Nature}\ }\textbf {\bibinfo {volume} {621}},\ \bibinfo {pages}
  {728} (\bibinfo {year} {2023})}\BibitemShut {NoStop}%
\bibitem [{\citenamefont {Sandvik}(2003)}]{Sandvik2003}%
  \BibitemOpen
  \bibfield  {author} {\bibinfo {author} {\bibfnamefont {A.~W.}\ \bibnamefont
  {Sandvik}},\ }\bibfield  {title} {\bibinfo {title} {Stochastic series
  expansion method for quantum {I}sing models with arbitrary interactions},\
  }\href {https://doi.org/10.1103/physreve.68.056701} {\bibfield  {journal}
  {\bibinfo  {journal} {Phys. Rev. E}\ }\textbf {\bibinfo {volume} {68}},\
  \bibinfo {pages} {056701} (\bibinfo {year} {2003})}\BibitemShut {NoStop}%
\bibitem [{\citenamefont {Troyer}\ and\ \citenamefont
  {Wiese}(2005)}]{TroyerWiese2005}%
  \BibitemOpen
  \bibfield  {author} {\bibinfo {author} {\bibfnamefont {M.}~\bibnamefont
  {Troyer}}\ and\ \bibinfo {author} {\bibfnamefont {U.-J.}\ \bibnamefont
  {Wiese}},\ }\bibfield  {title} {\bibinfo {title} {Computational complexity
  and fundamental limitations to fermionic quantum {M}onte {C}arlo
  simulations},\ }\href {https://doi.org/10.1103/physrevlett.94.170201}
  {\bibfield  {journal} {\bibinfo  {journal} {Phys. Rev. Lett.}\ }\textbf
  {\bibinfo {volume} {94}},\ \bibinfo {pages} {170201} (\bibinfo {year}
  {2005})}\BibitemShut {NoStop}%
\bibitem [{\citenamefont {Sbierski}\ \emph {et~al.}(2024)\citenamefont
  {Sbierski}, \citenamefont {Bintz}, \citenamefont {Chatterjee}, \citenamefont
  {Schuler}, \citenamefont {Yao},\ and\ \citenamefont {Pollet}}]{Sbierski2024}%
  \BibitemOpen
  \bibfield  {author} {\bibinfo {author} {\bibfnamefont {B.}~\bibnamefont
  {Sbierski}}, \bibinfo {author} {\bibfnamefont {M.}~\bibnamefont {Bintz}},
  \bibinfo {author} {\bibfnamefont {S.}~\bibnamefont {Chatterjee}}, \bibinfo
  {author} {\bibfnamefont {M.}~\bibnamefont {Schuler}}, \bibinfo {author}
  {\bibfnamefont {N.~Y.}\ \bibnamefont {Yao}},\ and\ \bibinfo {author}
  {\bibfnamefont {L.}~\bibnamefont {Pollet}},\ }\bibfield  {title} {\bibinfo
  {title} {Magnetism in the two-dimensional dipolar {XY} model},\ }\href
  {https://doi.org/10.1103/PhysRevB.109.144411} {\bibfield  {journal} {\bibinfo
   {journal} {Phys. Rev. B}\ }\textbf {\bibinfo {volume} {109}},\ \bibinfo
  {pages} {144411} (\bibinfo {year} {2024})}\BibitemShut {NoStop}%
\bibitem [{\citenamefont {Crosswhite}\ \emph {et~al.}(2008)\citenamefont
  {Crosswhite}, \citenamefont {Doherty},\ and\ \citenamefont
  {Vidal}}]{Crosswhite2008}%
  \BibitemOpen
  \bibfield  {author} {\bibinfo {author} {\bibfnamefont {G.~M.}\ \bibnamefont
  {Crosswhite}}, \bibinfo {author} {\bibfnamefont {A.~C.}\ \bibnamefont
  {Doherty}},\ and\ \bibinfo {author} {\bibfnamefont {G.}~\bibnamefont
  {Vidal}},\ }\bibfield  {title} {\bibinfo {title} {Applying matrix product
  operators to model systems with long-range interactions},\ }\href
  {https://doi.org/10.1103/physrevb.78.035116} {\bibfield  {journal} {\bibinfo
  {journal} {Phys. Rev. B}\ }\textbf {\bibinfo {volume} {78}},\ \bibinfo
  {pages} {035116} (\bibinfo {year} {2008})}\BibitemShut {NoStop}%
\bibitem [{\citenamefont {Zaletel}\ \emph {et~al.}(2015)\citenamefont
  {Zaletel}, \citenamefont {Mong}, \citenamefont {Karrasch}, \citenamefont
  {Moore},\ and\ \citenamefont {Pollmann}}]{Zaletel2015}%
  \BibitemOpen
  \bibfield  {author} {\bibinfo {author} {\bibfnamefont {M.~P.}\ \bibnamefont
  {Zaletel}}, \bibinfo {author} {\bibfnamefont {R.~S.~K.}\ \bibnamefont
  {Mong}}, \bibinfo {author} {\bibfnamefont {C.}~\bibnamefont {Karrasch}},
  \bibinfo {author} {\bibfnamefont {J.~E.}\ \bibnamefont {Moore}},\ and\
  \bibinfo {author} {\bibfnamefont {F.}~\bibnamefont {Pollmann}},\ }\bibfield
  {title} {\bibinfo {title} {Time-evolving a matrix product state with
  long-ranged interactions},\ }\href
  {https://doi.org/10.1103/physrevb.91.165112} {\bibfield  {journal} {\bibinfo
  {journal} {Phys. Rev. B}\ }\textbf {\bibinfo {volume} {91}},\ \bibinfo
  {pages} {165112} (\bibinfo {year} {2015})}\BibitemShut {NoStop}%
\bibitem [{\citenamefont {Stoudenmire}\ and\ \citenamefont
  {White}(2012)}]{StoudenmireWhite2012}%
  \BibitemOpen
  \bibfield  {author} {\bibinfo {author} {\bibfnamefont {E.}~\bibnamefont
  {Stoudenmire}}\ and\ \bibinfo {author} {\bibfnamefont {S.~R.}\ \bibnamefont
  {White}},\ }\bibfield  {title} {\bibinfo {title} {Studying two-dimensional
  systems with the density matrix renormalization group},\ }\href
  {https://doi.org/10.1146/annurev-conmatphys-020911-125018} {\bibfield
  {journal} {\bibinfo  {journal} {Annu. Rev. Condens. Matter Phys.}\ }\textbf
  {\bibinfo {volume} {3}},\ \bibinfo {pages} {111} (\bibinfo {year}
  {2012})}\BibitemShut {NoStop}%
\bibitem [{\citenamefont {Xiang}(2023)}]{xiang2023density}%
  \BibitemOpen
  \bibfield  {author} {\bibinfo {author} {\bibfnamefont {T.}~\bibnamefont
  {Xiang}},\ }\href
  {https://books.google.com/books?hl=en&lr=&id=E5fxEAAAQBAJ&oi=fnd&pg=PP1&ots=Hqdq-ApAx0&sig=xi-IvwDkPLk7CTWPcB_d5zRtFZk}
  {\emph {\bibinfo {title} {{Density Matrix} and {Tensor Network
  Renormalization}}}}\ (\bibinfo  {publisher} {Cambridge University Press},\
  \bibinfo {year} {2023})\BibitemShut {NoStop}%
\bibitem [{\citenamefont {Jiang}\ \emph {et~al.}(2008)\citenamefont {Jiang},
  \citenamefont {Weng},\ and\ \citenamefont
  {Xiang}}]{jiangAccurateDeterminationTensor2008b}%
  \BibitemOpen
  \bibfield  {author} {\bibinfo {author} {\bibfnamefont {H.~C.}\ \bibnamefont
  {Jiang}}, \bibinfo {author} {\bibfnamefont {Z.~Y.}\ \bibnamefont {Weng}},\
  and\ \bibinfo {author} {\bibfnamefont {T.}~\bibnamefont {Xiang}},\ }\bibfield
   {title} {\bibinfo {title} {Accurate {{Determination}} of {{Tensor Network
  State}} of {{Quantum Lattice Models}} in {{Two Dimensions}}},\ }\href
  {https://doi.org/10.1103/PhysRevLett.101.090603} {\bibfield  {journal}
  {\bibinfo  {journal} {Phys. Rev. Lett.}\ }\textbf {\bibinfo {volume} {101}},\
  \bibinfo {pages} {090603} (\bibinfo {year} {2008})}\BibitemShut {NoStop}%
\bibitem [{\citenamefont {Liu}\ \emph {et~al.}(2021)\citenamefont {Liu},
  \citenamefont {Huang}, \citenamefont {Gong},\ and\ \citenamefont
  {Gu}}]{Liu2021}%
  \BibitemOpen
  \bibfield  {author} {\bibinfo {author} {\bibfnamefont {W.-Y.}\ \bibnamefont
  {Liu}}, \bibinfo {author} {\bibfnamefont {Y.-Z.}\ \bibnamefont {Huang}},
  \bibinfo {author} {\bibfnamefont {S.-S.}\ \bibnamefont {Gong}},\ and\
  \bibinfo {author} {\bibfnamefont {Z.-C.}\ \bibnamefont {Gu}},\ }\bibfield
  {title} {\bibinfo {title} {Accurate simulation for finite projected entangled
  pair states in two dimensions},\ }\href
  {https://doi.org/10.1103/PhysRevB.103.235155} {\bibfield  {journal} {\bibinfo
   {journal} {Phys. Rev. B}\ }\textbf {\bibinfo {volume} {103}},\ \bibinfo
  {pages} {235155} (\bibinfo {year} {2021})}\BibitemShut {NoStop}%
\bibitem [{\citenamefont {Vieijra}\ \emph {et~al.}(2021)\citenamefont
  {Vieijra}, \citenamefont {Haegeman}, \citenamefont {Verstraete},\ and\
  \citenamefont {Vanderstraeten}}]{Vieijra2021}%
  \BibitemOpen
  \bibfield  {author} {\bibinfo {author} {\bibfnamefont {T.}~\bibnamefont
  {Vieijra}}, \bibinfo {author} {\bibfnamefont {J.}~\bibnamefont {Haegeman}},
  \bibinfo {author} {\bibfnamefont {F.}~\bibnamefont {Verstraete}},\ and\
  \bibinfo {author} {\bibfnamefont {L.}~\bibnamefont {Vanderstraeten}},\
  }\bibfield  {title} {\bibinfo {title} {Direct sampling of projected
  entangled-pair states},\ }\href {https://doi.org/10.1103/PhysRevB.104.235141}
  {\bibfield  {journal} {\bibinfo  {journal} {Phys. Rev. B}\ }\textbf {\bibinfo
  {volume} {104}},\ \bibinfo {pages} {235141} (\bibinfo {year}
  {2021})}\BibitemShut {NoStop}%
\bibitem [{\citenamefont {Wu}\ and\ \citenamefont {Dai}(2026)}]{WuDai2026}%
  \BibitemOpen
  \bibfield  {author} {\bibinfo {author} {\bibfnamefont {Y.}~\bibnamefont
  {Wu}}\ and\ \bibinfo {author} {\bibfnamefont {Z.}~\bibnamefont {Dai}},\
  }\bibfield  {title} {\bibinfo {title} {Algorithms for variational {M}onte
  {C}arlo calculations of fermion projected entangled pair states in the swap
  gates formulation and the detailed balance of tensor network sequential
  sampling},\ }\href {https://doi.org/10.1088/1674-1056/ae2673} {\bibfield
  {journal} {\bibinfo  {journal} {Chin. Phys. B}\ }\textbf {\bibinfo {volume}
  {35}},\ \bibinfo {pages} {020502} (\bibinfo {year} {2026})}\BibitemShut
  {NoStop}%
\bibitem [{\citenamefont {Chen}\ \emph
  {et~al.}(2026{\natexlab{a}})\citenamefont {Chen}, \citenamefont {Liu},
  \citenamefont {Wu}, \citenamefont {Zhang},\ and\ \citenamefont
  {Deng}}]{chen2026variationalmontecarlovmc}%
  \BibitemOpen
  \bibfield  {author} {\bibinfo {author} {\bibfnamefont {T.}~\bibnamefont
  {Chen}}, \bibinfo {author} {\bibfnamefont {J.}~\bibnamefont {Liu}}, \bibinfo
  {author} {\bibfnamefont {Y.}~\bibnamefont {Wu}}, \bibinfo {author}
  {\bibfnamefont {P.}~\bibnamefont {Zhang}},\ and\ \bibinfo {author}
  {\bibfnamefont {Y.}~\bibnamefont {Deng}},\ }\bibfield  {title} {\bibinfo
  {title} {Variational {M}onte {C}arlo with row-update projected entangled-pair
  states and its applications to quantum spin glasses},\ }\href
  {https://doi.org/10.1103/9z82-vlky} {\bibfield  {journal} {\bibinfo
  {journal} {Phys. Rev. B}\ }\textbf {\bibinfo {volume} {114}},\ \bibinfo
  {pages} {144201} (\bibinfo {year} {2026}{\natexlab{a}})}\BibitemShut
  {NoStop}%
\bibitem [{\citenamefont {Carleo}\ and\ \citenamefont
  {Troyer}(2017)}]{CarleoTroyer2017}%
  \BibitemOpen
  \bibfield  {author} {\bibinfo {author} {\bibfnamefont {G.}~\bibnamefont
  {Carleo}}\ and\ \bibinfo {author} {\bibfnamefont {M.}~\bibnamefont
  {Troyer}},\ }\bibfield  {title} {\bibinfo {title} {Solving the quantum
  many-body problem with artificial neural networks},\ }\href
  {https://doi.org/10.1126/science.aag2302} {\bibfield  {journal} {\bibinfo
  {journal} {Science}\ }\textbf {\bibinfo {volume} {355}},\ \bibinfo {pages}
  {602} (\bibinfo {year} {2017})}\BibitemShut {NoStop}%
\bibitem [{\citenamefont {Sharir}\ \emph {et~al.}(2020)\citenamefont {Sharir},
  \citenamefont {Levine}, \citenamefont {Wies}, \citenamefont {Carleo},\ and\
  \citenamefont {Shashua}}]{Sharir2020}%
  \BibitemOpen
  \bibfield  {author} {\bibinfo {author} {\bibfnamefont {O.}~\bibnamefont
  {Sharir}}, \bibinfo {author} {\bibfnamefont {Y.}~\bibnamefont {Levine}},
  \bibinfo {author} {\bibfnamefont {N.}~\bibnamefont {Wies}}, \bibinfo {author}
  {\bibfnamefont {G.}~\bibnamefont {Carleo}},\ and\ \bibinfo {author}
  {\bibfnamefont {A.}~\bibnamefont {Shashua}},\ }\bibfield  {title} {\bibinfo
  {title} {Deep autoregressive models for the efficient variational simulation
  of many-body quantum systems},\ }\href
  {https://doi.org/10.1103/PhysRevLett.124.020503} {\bibfield  {journal}
  {\bibinfo  {journal} {Phys. Rev. Lett.}\ }\textbf {\bibinfo {volume} {124}},\
  \bibinfo {pages} {020503} (\bibinfo {year} {2020})}\BibitemShut {NoStop}%
\bibitem [{\citenamefont {Sorella}(2001)}]{Sorella2001}%
  \BibitemOpen
  \bibfield  {author} {\bibinfo {author} {\bibfnamefont {S.}~\bibnamefont
  {Sorella}},\ }\bibfield  {title} {\bibinfo {title} {Generalized {L}anczos
  algorithm for variational quantum {M}onte {C}arlo},\ }\href
  {https://doi.org/10.1103/PhysRevB.64.024512} {\bibfield  {journal} {\bibinfo
  {journal} {Phys. Rev. B}\ }\textbf {\bibinfo {volume} {64}},\ \bibinfo
  {pages} {024512} (\bibinfo {year} {2001})}\BibitemShut {NoStop}%
\bibitem [{\citenamefont {Carleo}\ \emph {et~al.}(2017)\citenamefont {Carleo},
  \citenamefont {Cevolani}, \citenamefont {Sanchez-Palencia},\ and\
  \citenamefont {Holzmann}}]{Carleo2017}%
  \BibitemOpen
  \bibfield  {author} {\bibinfo {author} {\bibfnamefont {G.}~\bibnamefont
  {Carleo}}, \bibinfo {author} {\bibfnamefont {L.}~\bibnamefont {Cevolani}},
  \bibinfo {author} {\bibfnamefont {L.}~\bibnamefont {Sanchez-Palencia}},\ and\
  \bibinfo {author} {\bibfnamefont {M.}~\bibnamefont {Holzmann}},\ }\bibfield
  {title} {\bibinfo {title} {Unitary dynamics of strongly interacting {B}ose
  gases with the time-dependent variational {M}onte {C}arlo method in
  continuous space},\ }\href {https://doi.org/10.1103/PhysRevX.7.031026}
  {\bibfield  {journal} {\bibinfo  {journal} {Phys. Rev. X}\ }\textbf {\bibinfo
  {volume} {7}},\ \bibinfo {pages} {031026} (\bibinfo {year}
  {2017})}\BibitemShut {NoStop}%
\bibitem [{\citenamefont {Schmitt}\ and\ \citenamefont
  {Heyl}(2020)}]{SchmittHeyl2020}%
  \BibitemOpen
  \bibfield  {author} {\bibinfo {author} {\bibfnamefont {M.}~\bibnamefont
  {Schmitt}}\ and\ \bibinfo {author} {\bibfnamefont {M.}~\bibnamefont {Heyl}},\
  }\bibfield  {title} {\bibinfo {title} {Quantum many-body dynamics in two
  dimensions with artificial neural networks},\ }\href
  {https://doi.org/10.1103/PhysRevLett.125.100503} {\bibfield  {journal}
  {\bibinfo  {journal} {Phys. Rev. Lett.}\ }\textbf {\bibinfo {volume} {125}},\
  \bibinfo {pages} {100503} (\bibinfo {year} {2020})}\BibitemShut {NoStop}%
\bibitem [{\citenamefont {Comparin}\ \emph {et~al.}(2022)\citenamefont
  {Comparin}, \citenamefont {Mezzacapo},\ and\ \citenamefont
  {Roscilde}}]{Comparin2022}%
  \BibitemOpen
  \bibfield  {author} {\bibinfo {author} {\bibfnamefont {T.}~\bibnamefont
  {Comparin}}, \bibinfo {author} {\bibfnamefont {F.}~\bibnamefont
  {Mezzacapo}},\ and\ \bibinfo {author} {\bibfnamefont {T.}~\bibnamefont
  {Roscilde}},\ }\bibfield  {title} {\bibinfo {title} {Multipartite entangled
  states in dipolar quantum simulators},\ }\href
  {https://doi.org/10.1103/PhysRevLett.129.150503} {\bibfield  {journal}
  {\bibinfo  {journal} {Phys. Rev. Lett.}\ }\textbf {\bibinfo {volume} {129}},\
  \bibinfo {pages} {150503} (\bibinfo {year} {2022})}\BibitemShut {NoStop}%
\bibitem [{\citenamefont {Wu}\ and\ \citenamefont {Nys}(2026)}]{Wu2026}%
  \BibitemOpen
  \bibfield  {author} {\bibinfo {author} {\bibfnamefont {Y.}~\bibnamefont
  {Wu}}\ and\ \bibinfo {author} {\bibfnamefont {J.}~\bibnamefont {Nys}},\
  }\bibfield  {title} {\bibinfo {title} {Real-time dynamics in two dimensions
  with tensor network states via time-dependent variational {M}onte {C}arlo
  method},\ }\href {https://doi.org/10.1103/tggc-8fjx} {\bibfield  {journal}
  {\bibinfo  {journal} {PRX Quantum}\ }\textbf {\bibinfo {volume} {7}},\
  \bibinfo {pages} {033035} (\bibinfo {year} {2026})}\BibitemShut {NoStop}%
\bibitem [{\citenamefont {Mendes-Santos}\ \emph {et~al.}(2023)\citenamefont
  {Mendes-Santos}, \citenamefont {Schmitt},\ and\ \citenamefont
  {Heyl}}]{heyl_prl}%
  \BibitemOpen
  \bibfield  {author} {\bibinfo {author} {\bibfnamefont {T.}~\bibnamefont
  {Mendes-Santos}}, \bibinfo {author} {\bibfnamefont {M.}~\bibnamefont
  {Schmitt}},\ and\ \bibinfo {author} {\bibfnamefont {M.}~\bibnamefont
  {Heyl}},\ }\bibfield  {title} {\bibinfo {title} {Highly resolved spectral
  functions of two-dimensional systems with neural quantum states},\ }\href
  {https://doi.org/10.1103/PhysRevLett.131.046501} {\bibfield  {journal}
  {\bibinfo  {journal} {Phys. Rev. Lett.}\ }\textbf {\bibinfo {volume} {131}},\
  \bibinfo {pages} {046501} (\bibinfo {year} {2023})}\BibitemShut {NoStop}%
\bibitem [{\citenamefont {Mauron}\ \emph {et~al.}(2025)\citenamefont {Mauron},
  \citenamefont {Denis}, \citenamefont {Nys},\ and\ \citenamefont
  {Carleo}}]{jannes_np}%
  \BibitemOpen
  \bibfield  {author} {\bibinfo {author} {\bibfnamefont {L.}~\bibnamefont
  {Mauron}}, \bibinfo {author} {\bibfnamefont {Z.}~\bibnamefont {Denis}},
  \bibinfo {author} {\bibfnamefont {J.}~\bibnamefont {Nys}},\ and\ \bibinfo
  {author} {\bibfnamefont {G.}~\bibnamefont {Carleo}},\ }\bibfield  {title}
  {\bibinfo {title} {Predicting topological entanglement entropy in a {R}ydberg
  analogue simulator},\ }\href {https://doi.org/10.1038/s41567-025-02944-3}
  {\bibfield  {journal} {\bibinfo  {journal} {Nat. Phys.}\ }\textbf {\bibinfo
  {volume} {21}},\ \bibinfo {pages} {1332} (\bibinfo {year}
  {2025})}\BibitemShut {NoStop}%
\bibitem [{\citenamefont {Schr{\"o}der}\ \emph {et~al.}(2026)\citenamefont
  {Schr{\"o}der}, \citenamefont {Weerda},\ and\ \citenamefont
  {Rizzi}}]{rizzi_arxiv}%
  \BibitemOpen
  \bibfield  {author} {\bibinfo {author} {\bibfnamefont {K.}~\bibnamefont
  {Schr{\"o}der}}, \bibinfo {author} {\bibfnamefont {E.~L.}\ \bibnamefont
  {Weerda}},\ and\ \bibinfo {author} {\bibfnamefont {M.}~\bibnamefont
  {Rizzi}},\ }\href@noop {} {\bibinfo {title} {Exploring ground state phase
  diagrams of {R}ydberg atom arrays with finite projected entangled-pair
  states}} (\bibinfo {year} {2026}),\ \Eprint
  {https://arxiv.org/abs/2610.26520} {arXiv:2610.26520 [cond-mat.quant-gas]}
  \BibitemShut {NoStop}%
\bibitem [{\citenamefont {Haegeman}\ \emph {et~al.}(2011)\citenamefont
  {Haegeman}, \citenamefont {Cirac}, \citenamefont {Osborne}, \citenamefont
  {Pi{\v z}orn}, \citenamefont {Verschelde},\ and\ \citenamefont
  {Verstraete}}]{Haegeman2011}%
  \BibitemOpen
  \bibfield  {author} {\bibinfo {author} {\bibfnamefont {J.}~\bibnamefont
  {Haegeman}}, \bibinfo {author} {\bibfnamefont {J.~I.}\ \bibnamefont {Cirac}},
  \bibinfo {author} {\bibfnamefont {T.~J.}\ \bibnamefont {Osborne}}, \bibinfo
  {author} {\bibfnamefont {I.}~\bibnamefont {Pi{\v z}orn}}, \bibinfo {author}
  {\bibfnamefont {H.}~\bibnamefont {Verschelde}},\ and\ \bibinfo {author}
  {\bibfnamefont {F.}~\bibnamefont {Verstraete}},\ }\bibfield  {title}
  {\bibinfo {title} {Time-dependent variational principle for quantum
  lattices},\ }\href {https://doi.org/10.1103/PhysRevLett.107.070601}
  {\bibfield  {journal} {\bibinfo  {journal} {Phys. Rev. Lett.}\ }\textbf
  {\bibinfo {volume} {107}},\ \bibinfo {pages} {070601} (\bibinfo {year}
  {2011})}\BibitemShut {NoStop}%
\bibitem [{\citenamefont {Carleo}\ \emph {et~al.}(2012)\citenamefont {Carleo},
  \citenamefont {Becca}, \citenamefont {Schir{\'o}},\ and\ \citenamefont
  {Fabrizio}}]{Carleo2012}%
  \BibitemOpen
  \bibfield  {author} {\bibinfo {author} {\bibfnamefont {G.}~\bibnamefont
  {Carleo}}, \bibinfo {author} {\bibfnamefont {F.}~\bibnamefont {Becca}},
  \bibinfo {author} {\bibfnamefont {M.}~\bibnamefont {Schir{\'o}}},\ and\
  \bibinfo {author} {\bibfnamefont {M.}~\bibnamefont {Fabrizio}},\ }\bibfield
  {title} {\bibinfo {title} {Localization and glassy dynamics of many-body
  quantum systems},\ }\href {https://doi.org/10.1038/srep00243} {\bibfield
  {journal} {\bibinfo  {journal} {Sci. Rep.}\ }\textbf {\bibinfo {volume}
  {2}},\ \bibinfo {pages} {243} (\bibinfo {year} {2012})}\BibitemShut {NoStop}%
\bibitem [{\citenamefont {Provost}\ and\ \citenamefont
  {Vallee}(1980)}]{ProvostVallee1980}%
  \BibitemOpen
  \bibfield  {author} {\bibinfo {author} {\bibfnamefont {J.~P.}\ \bibnamefont
  {Provost}}\ and\ \bibinfo {author} {\bibfnamefont {G.}~\bibnamefont
  {Vallee}},\ }\bibfield  {title} {\bibinfo {title} {Riemannian structure on
  manifolds of quantum states},\ }\href {https://doi.org/10.1007/bf02193559}
  {\bibfield  {journal} {\bibinfo  {journal} {Commun. Math. Phys.}\ }\textbf
  {\bibinfo {volume} {76}},\ \bibinfo {pages} {289} (\bibinfo {year}
  {1980})}\BibitemShut {NoStop}%
\bibitem [{\citenamefont {Assaraf}\ and\ \citenamefont
  {Caffarel}(1999)}]{AssarafCaffarel1999}%
  \BibitemOpen
  \bibfield  {author} {\bibinfo {author} {\bibfnamefont {R.}~\bibnamefont
  {Assaraf}}\ and\ \bibinfo {author} {\bibfnamefont {M.}~\bibnamefont
  {Caffarel}},\ }\bibfield  {title} {\bibinfo {title} {Zero-variance principle
  for {M}onte {C}arlo algorithms},\ }\href
  {https://doi.org/10.1103/PhysRevLett.83.4682} {\bibfield  {journal} {\bibinfo
   {journal} {Phys. Rev. Lett.}\ }\textbf {\bibinfo {volume} {83}},\ \bibinfo
  {pages} {4682} (\bibinfo {year} {1999})}\BibitemShut {NoStop}%
\bibitem [{\citenamefont {Sandvik}\ and\ \citenamefont
  {Hamer}(1999)}]{Sandvik1999}%
  \BibitemOpen
  \bibfield  {author} {\bibinfo {author} {\bibfnamefont {A.~W.}\ \bibnamefont
  {Sandvik}}\ and\ \bibinfo {author} {\bibfnamefont {C.~J.}\ \bibnamefont
  {Hamer}},\ }\bibfield  {title} {\bibinfo {title} {Ground-state parameters,
  finite-size scaling, and low-temperature properties of the two-dimensional
  {$S=1/2$} {XY} model},\ }\href {https://doi.org/10.1103/PhysRevB.60.6588}
  {\bibfield  {journal} {\bibinfo  {journal} {Phys. Rev. B}\ }\textbf {\bibinfo
  {volume} {60}},\ \bibinfo {pages} {6588} (\bibinfo {year}
  {1999})}\BibitemShut {NoStop}%
\bibitem [{\citenamefont {Signoles}\ \emph {et~al.}(2021)\citenamefont
  {Signoles}, \citenamefont {Franz}, \citenamefont {Ferracini~Alves},
  \citenamefont {G{\"a}rttner}, \citenamefont {Whitlock}, \citenamefont
  {Z{\"u}rn},\ and\ \citenamefont {Weidem{\"u}ller}}]{Signoles2021}%
  \BibitemOpen
  \bibfield  {author} {\bibinfo {author} {\bibfnamefont {A.}~\bibnamefont
  {Signoles}}, \bibinfo {author} {\bibfnamefont {T.}~\bibnamefont {Franz}},
  \bibinfo {author} {\bibfnamefont {R.}~\bibnamefont {Ferracini~Alves}},
  \bibinfo {author} {\bibfnamefont {M.}~\bibnamefont {G{\"a}rttner}}, \bibinfo
  {author} {\bibfnamefont {S.}~\bibnamefont {Whitlock}}, \bibinfo {author}
  {\bibfnamefont {G.}~\bibnamefont {Z{\"u}rn}},\ and\ \bibinfo {author}
  {\bibfnamefont {M.}~\bibnamefont {Weidem{\"u}ller}},\ }\bibfield  {title}
  {\bibinfo {title} {Glassy dynamics in a disordered {H}eisenberg quantum spin
  system},\ }\href {https://doi.org/10.1103/PhysRevX.11.011011} {\bibfield
  {journal} {\bibinfo  {journal} {Phys. Rev. X}\ }\textbf {\bibinfo {volume}
  {11}},\ \bibinfo {pages} {011011} (\bibinfo {year} {2021})}\BibitemShut
  {NoStop}%
\bibitem [{\citenamefont {Islam}\ \emph {et~al.}(2013)\citenamefont {Islam},
  \citenamefont {Senko}, \citenamefont {Campbell}, \citenamefont {Korenblit},
  \citenamefont {Smith}, \citenamefont {Lee}, \citenamefont {Edwards},
  \citenamefont {Wang}, \citenamefont {Freericks},\ and\ \citenamefont
  {Monroe}}]{Islam2013}%
  \BibitemOpen
  \bibfield  {author} {\bibinfo {author} {\bibfnamefont {R.}~\bibnamefont
  {Islam}}, \bibinfo {author} {\bibfnamefont {C.}~\bibnamefont {Senko}},
  \bibinfo {author} {\bibfnamefont {W.~C.}\ \bibnamefont {Campbell}}, \bibinfo
  {author} {\bibfnamefont {S.}~\bibnamefont {Korenblit}}, \bibinfo {author}
  {\bibfnamefont {J.}~\bibnamefont {Smith}}, \bibinfo {author} {\bibfnamefont
  {A.}~\bibnamefont {Lee}}, \bibinfo {author} {\bibfnamefont {E.~E.}\
  \bibnamefont {Edwards}}, \bibinfo {author} {\bibfnamefont {C.-C.~J.}\
  \bibnamefont {Wang}}, \bibinfo {author} {\bibfnamefont {J.~K.}\ \bibnamefont
  {Freericks}},\ and\ \bibinfo {author} {\bibfnamefont {C.}~\bibnamefont
  {Monroe}},\ }\bibfield  {title} {\bibinfo {title} {Emergence and frustration
  of magnetism with variable-range interactions in a quantum simulator},\
  }\href {https://doi.org/10.1126/science.1232296} {\bibfield  {journal}
  {\bibinfo  {journal} {Science}\ }\textbf {\bibinfo {volume} {340}},\ \bibinfo
  {pages} {583} (\bibinfo {year} {2013})}\BibitemShut {NoStop}%
\bibitem [{\citenamefont {Kennedy}\ \emph {et~al.}(1988)\citenamefont
  {Kennedy}, \citenamefont {Lieb},\ and\ \citenamefont
  {Shastry}}]{Kennedy1988}%
  \BibitemOpen
  \bibfield  {author} {\bibinfo {author} {\bibfnamefont {T.}~\bibnamefont
  {Kennedy}}, \bibinfo {author} {\bibfnamefont {E.~H.}\ \bibnamefont {Lieb}},\
  and\ \bibinfo {author} {\bibfnamefont {B.~S.}\ \bibnamefont {Shastry}},\
  }\bibfield  {title} {\bibinfo {title} {The {XY} model has long-range order
  for all spins and all dimensions greater than one},\ }\href
  {https://doi.org/10.1103/PhysRevLett.61.2582} {\bibfield  {journal} {\bibinfo
   {journal} {Phys. Rev. Lett.}\ }\textbf {\bibinfo {volume} {61}},\ \bibinfo
  {pages} {2582} (\bibinfo {year} {1988})}\BibitemShut {NoStop}%
\bibitem [{\citenamefont {Kubo}(1988)}]{Kubo1988}%
  \BibitemOpen
  \bibfield  {author} {\bibinfo {author} {\bibfnamefont {K.}~\bibnamefont
  {Kubo}},\ }\bibfield  {title} {\bibinfo {title} {Existence of long-range
  order in the {XY} model},\ }\href
  {https://doi.org/10.1103/PhysRevLett.61.110} {\bibfield  {journal} {\bibinfo
  {journal} {Phys. Rev. Lett.}\ }\textbf {\bibinfo {volume} {61}},\ \bibinfo
  {pages} {110} (\bibinfo {year} {1988})}\BibitemShut {NoStop}%
\bibitem [{\citenamefont {Sandvik}\ and\ \citenamefont
  {Kurkij{\"a}rvi}(1991)}]{SandvikKurkijarvi1991}%
  \BibitemOpen
  \bibfield  {author} {\bibinfo {author} {\bibfnamefont {A.~W.}\ \bibnamefont
  {Sandvik}}\ and\ \bibinfo {author} {\bibfnamefont {J.}~\bibnamefont
  {Kurkij{\"a}rvi}},\ }\bibfield  {title} {\bibinfo {title} {Quantum {Monte
  Carlo} simulation method for spin systems},\ }\href
  {https://doi.org/10.1103/PhysRevB.43.5950} {\bibfield  {journal} {\bibinfo
  {journal} {Phys. Rev. B}\ }\textbf {\bibinfo {volume} {43}},\ \bibinfo
  {pages} {5950} (\bibinfo {year} {1991})}\BibitemShut {NoStop}%
\bibitem [{\citenamefont {Sylju{\aa}sen}\ and\ \citenamefont
  {Sandvik}(2002)}]{SyljuasenSandvik2002}%
  \BibitemOpen
  \bibfield  {author} {\bibinfo {author} {\bibfnamefont {O.~F.}\ \bibnamefont
  {Sylju{\aa}sen}}\ and\ \bibinfo {author} {\bibfnamefont {A.~W.}\ \bibnamefont
  {Sandvik}},\ }\bibfield  {title} {\bibinfo {title} {Quantum {Monte Carlo}
  with directed loops},\ }\href {https://doi.org/10.1103/PhysRevE.66.046701}
  {\bibfield  {journal} {\bibinfo  {journal} {Phys. Rev. E}\ }\textbf {\bibinfo
  {volume} {66}},\ \bibinfo {pages} {046701} (\bibinfo {year}
  {2002})}\BibitemShut {NoStop}%
\bibitem [{\citenamefont {White}\ and\ \citenamefont
  {Chernyshev}(2007)}]{WhiteChernyshev2007}%
  \BibitemOpen
  \bibfield  {author} {\bibinfo {author} {\bibfnamefont {S.~R.}\ \bibnamefont
  {White}}\ and\ \bibinfo {author} {\bibfnamefont {A.~L.}\ \bibnamefont
  {Chernyshev}},\ }\bibfield  {title} {\bibinfo {title} {N{\'e}el order in
  square and triangular lattice {H}eisenberg models},\ }\href
  {https://doi.org/10.1103/PhysRevLett.99.127004} {\bibfield  {journal}
  {\bibinfo  {journal} {Phys. Rev. Lett.}\ }\textbf {\bibinfo {volume} {99}},\
  \bibinfo {pages} {127004} (\bibinfo {year} {2007})}\BibitemShut {NoStop}%
\bibitem [{\citenamefont {Assaad}\ and\ \citenamefont
  {Herbut}(2013)}]{AssaadHerbut2013}%
  \BibitemOpen
  \bibfield  {author} {\bibinfo {author} {\bibfnamefont {F.~F.}\ \bibnamefont
  {Assaad}}\ and\ \bibinfo {author} {\bibfnamefont {I.~F.}\ \bibnamefont
  {Herbut}},\ }\bibfield  {title} {\bibinfo {title} {Pinning the order: The
  nature of quantum criticality in the {H}ubbard model on honeycomb lattice},\
  }\href {https://doi.org/10.1103/PhysRevX.3.031010} {\bibfield  {journal}
  {\bibinfo  {journal} {Phys. Rev. X}\ }\textbf {\bibinfo {volume} {3}},\
  \bibinfo {pages} {031010} (\bibinfo {year} {2013})}\BibitemShut {NoStop}%
\bibitem [{\citenamefont {Chen}\ \emph
  {et~al.}(2026{\natexlab{b}})\citenamefont {Chen}, \citenamefont {Xiang},\
  and\ \citenamefont {Wu}}]{sm}%
  \BibitemOpen
  \bibfield  {author} {\bibinfo {author} {\bibfnamefont {J.-L.}\ \bibnamefont
  {Chen}}, \bibinfo {author} {\bibfnamefont {T.}~\bibnamefont {Xiang}},\ and\
  \bibinfo {author} {\bibfnamefont {Y.}~\bibnamefont {Wu}},\ }\href@noop {}
  {\bibinfo {title} {Supplemental material}} (\bibinfo {year}
  {2026}{\natexlab{b}})\BibitemShut {NoStop}%
\bibitem [{\citenamefont {Robertson}(1929)}]{Robertson1929}%
  \BibitemOpen
  \bibfield  {author} {\bibinfo {author} {\bibfnamefont {H.~P.}\ \bibnamefont
  {Robertson}},\ }\bibfield  {title} {\bibinfo {title} {The uncertainty
  principle},\ }\href {https://doi.org/10.1103/PhysRev.34.163} {\bibfield
  {journal} {\bibinfo  {journal} {Phys. Rev.}\ }\textbf {\bibinfo {volume}
  {34}},\ \bibinfo {pages} {163} (\bibinfo {year} {1929})}\BibitemShut
  {NoStop}%
\bibitem [{\citenamefont {White}(1992)}]{White1992}%
  \BibitemOpen
  \bibfield  {author} {\bibinfo {author} {\bibfnamefont {S.~R.}\ \bibnamefont
  {White}},\ }\bibfield  {title} {\bibinfo {title} {Density matrix formulation
  for quantum renormalization groups},\ }\href
  {https://doi.org/10.1103/physrevlett.69.2863} {\bibfield  {journal} {\bibinfo
   {journal} {Phys. Rev. Lett.}\ }\textbf {\bibinfo {volume} {69}},\ \bibinfo
  {pages} {2863} (\bibinfo {year} {1992})}\BibitemShut {NoStop}%
\bibitem [{\citenamefont {Bintz}\ \emph {et~al.}(2026)\citenamefont {Bintz},
  \citenamefont {Khalifa}, \citenamefont {Liu}, \citenamefont {Hauschild},
  \citenamefont {Zaletel}, \citenamefont {Chatterjee},\ and\ \citenamefont
  {Yao}}]{Bintz2026}%
  \BibitemOpen
  \bibfield  {author} {\bibinfo {author} {\bibfnamefont {M.}~\bibnamefont
  {Bintz}}, \bibinfo {author} {\bibfnamefont {A.}~\bibnamefont {Khalifa}},
  \bibinfo {author} {\bibfnamefont {V.~S.}\ \bibnamefont {Liu}}, \bibinfo
  {author} {\bibfnamefont {J.}~\bibnamefont {Hauschild}}, \bibinfo {author}
  {\bibfnamefont {M.~P.}\ \bibnamefont {Zaletel}}, \bibinfo {author}
  {\bibfnamefont {S.}~\bibnamefont {Chatterjee}},\ and\ \bibinfo {author}
  {\bibfnamefont {N.~Y.}\ \bibnamefont {Yao}},\ }\href@noop {} {\bibinfo
  {title} {Ground states of quantum {XY} dipoles on the {A}rchimedean
  lattices}} (\bibinfo {year} {2026}),\ \Eprint
  {https://arxiv.org/abs/2605.07685} {arXiv:2605.07685 [cond-mat.str-el]}
  \BibitemShut {NoStop}%
\bibitem [{\citenamefont {Lubasch}\ \emph {et~al.}(2011)\citenamefont
  {Lubasch}, \citenamefont {Murg}, \citenamefont {Schneider}, \citenamefont
  {Cirac},\ and\ \citenamefont {Ba\~nuls}}]{Lubasch2011}%
  \BibitemOpen
  \bibfield  {author} {\bibinfo {author} {\bibfnamefont {M.}~\bibnamefont
  {Lubasch}}, \bibinfo {author} {\bibfnamefont {V.}~\bibnamefont {Murg}},
  \bibinfo {author} {\bibfnamefont {U.}~\bibnamefont {Schneider}}, \bibinfo
  {author} {\bibfnamefont {J.~I.}\ \bibnamefont {Cirac}},\ and\ \bibinfo
  {author} {\bibfnamefont {M.-C.}\ \bibnamefont {Ba\~nuls}},\ }\bibfield
  {title} {\bibinfo {title} {Adiabatic preparation of a {H}eisenberg
  antiferromagnet using an optical superlattice},\ }\href
  {https://doi.org/10.1103/PhysRevLett.107.165301} {\bibfield  {journal}
  {\bibinfo  {journal} {Phys. Rev. Lett.}\ }\textbf {\bibinfo {volume} {107}},\
  \bibinfo {pages} {165301} (\bibinfo {year} {2011})}\BibitemShut {NoStop}%
\bibitem [{\citenamefont {Kaneko}\ and\ \citenamefont
  {Danshita}(2022)}]{Kaneko2022}%
  \BibitemOpen
  \bibfield  {author} {\bibinfo {author} {\bibfnamefont {R.}~\bibnamefont
  {Kaneko}}\ and\ \bibinfo {author} {\bibfnamefont {I.}~\bibnamefont
  {Danshita}},\ }\bibfield  {title} {\bibinfo {title} {Tensor-network study of
  correlation-spreading dynamics in the two-dimensional {B}ose-{H}ubbard
  model},\ }\href {https://doi.org/10.1038/s42005-022-00848-9} {\bibfield
  {journal} {\bibinfo  {journal} {Commun. Phys.}\ }\textbf {\bibinfo {volume}
  {5}},\ \bibinfo {pages} {65} (\bibinfo {year} {2022})}\BibitemShut {NoStop}%
\bibitem [{\citenamefont {Liao}\ \emph {et~al.}(2023)\citenamefont {Liao},
  \citenamefont {Wang}, \citenamefont {Zhou}, \citenamefont {Zhang},\ and\
  \citenamefont {Xiang}}]{Liao2023}%
  \BibitemOpen
  \bibfield  {author} {\bibinfo {author} {\bibfnamefont {H.-J.}\ \bibnamefont
  {Liao}}, \bibinfo {author} {\bibfnamefont {K.}~\bibnamefont {Wang}}, \bibinfo
  {author} {\bibfnamefont {Z.-S.}\ \bibnamefont {Zhou}}, \bibinfo {author}
  {\bibfnamefont {P.}~\bibnamefont {Zhang}},\ and\ \bibinfo {author}
  {\bibfnamefont {T.}~\bibnamefont {Xiang}},\ }\href@noop {} {\bibinfo {title}
  {Simulation of {IBM}'s kicked {I}sing experiment with projected entangled
  pair operator}} (\bibinfo {year} {2023}),\ \Eprint
  {https://arxiv.org/abs/2308.03082} {arXiv:2308.03082 [quant-ph]} \BibitemShut
  {NoStop}%
\bibitem [{\citenamefont {Tindall}\ \emph {et~al.}(2024)\citenamefont
  {Tindall}, \citenamefont {Fishman}, \citenamefont {Stoudenmire},\ and\
  \citenamefont {Sels}}]{Tindall2024}%
  \BibitemOpen
  \bibfield  {author} {\bibinfo {author} {\bibfnamefont {J.}~\bibnamefont
  {Tindall}}, \bibinfo {author} {\bibfnamefont {M.}~\bibnamefont {Fishman}},
  \bibinfo {author} {\bibfnamefont {E.~M.}\ \bibnamefont {Stoudenmire}},\ and\
  \bibinfo {author} {\bibfnamefont {D.}~\bibnamefont {Sels}},\ }\bibfield
  {title} {\bibinfo {title} {Efficient tensor network simulation of {IBM}'s
  {E}agle kicked {I}sing experiment},\ }\href
  {https://doi.org/10.1103/PRXQuantum.5.010308} {\bibfield  {journal} {\bibinfo
   {journal} {PRX Quantum}\ }\textbf {\bibinfo {volume} {5}},\ \bibinfo {pages}
  {010308} (\bibinfo {year} {2024})}\BibitemShut {NoStop}%
\bibitem [{\citenamefont {Begu{\v{s}}i{\'c}}\ \emph {et~al.}(2024)\citenamefont
  {Begu{\v{s}}i{\'c}}, \citenamefont {Gray},\ and\ \citenamefont
  {Chan}}]{Begusic2024}%
  \BibitemOpen
  \bibfield  {author} {\bibinfo {author} {\bibfnamefont {T.}~\bibnamefont
  {Begu{\v{s}}i{\'c}}}, \bibinfo {author} {\bibfnamefont {J.}~\bibnamefont
  {Gray}},\ and\ \bibinfo {author} {\bibfnamefont {G.~K.-L.}\ \bibnamefont
  {Chan}},\ }\bibfield  {title} {\bibinfo {title} {Fast and converged classical
  simulations of evidence for the utility of quantum computing before fault
  tolerance},\ }\href {https://doi.org/10.1126/sciadv.adk4321} {\bibfield
  {journal} {\bibinfo  {journal} {Sci. Adv.}\ }\textbf {\bibinfo {volume}
  {10}},\ \bibinfo {pages} {eadk4321} (\bibinfo {year} {2024})}\BibitemShut
  {NoStop}%
\bibitem [{\citenamefont {Patra}\ \emph {et~al.}(2024)\citenamefont {Patra},
  \citenamefont {Jahromi}, \citenamefont {Singh},\ and\ \citenamefont
  {Or{\'u}s}}]{Patra2024}%
  \BibitemOpen
  \bibfield  {author} {\bibinfo {author} {\bibfnamefont {S.}~\bibnamefont
  {Patra}}, \bibinfo {author} {\bibfnamefont {S.~S.}\ \bibnamefont {Jahromi}},
  \bibinfo {author} {\bibfnamefont {S.}~\bibnamefont {Singh}},\ and\ \bibinfo
  {author} {\bibfnamefont {R.}~\bibnamefont {Or{\'u}s}},\ }\bibfield  {title}
  {\bibinfo {title} {Efficient tensor network simulation of {IBM}'s largest
  quantum processors},\ }\href
  {https://doi.org/10.1103/PhysRevResearch.6.013326} {\bibfield  {journal}
  {\bibinfo  {journal} {Phys. Rev. Res.}\ }\textbf {\bibinfo {volume} {6}},\
  \bibinfo {pages} {013326} (\bibinfo {year} {2024})}\BibitemShut {NoStop}%
\bibitem [{\citenamefont {Mauron}\ and\ \citenamefont
  {Carleo}(2025)}]{MauronCarleo2025}%
  \BibitemOpen
  \bibfield  {author} {\bibinfo {author} {\bibfnamefont {L.}~\bibnamefont
  {Mauron}}\ and\ \bibinfo {author} {\bibfnamefont {G.}~\bibnamefont
  {Carleo}},\ }\href@noop {} {\bibinfo {title} {Challenging the quantum
  advantage frontier with large-scale classical simulations of annealing
  dynamics}} (\bibinfo {year} {2025}),\ \Eprint
  {https://arxiv.org/abs/2503.08247} {arXiv:2503.08247 [quant-ph]} \BibitemShut
  {NoStop}%
\bibitem [{\citenamefont {Tindall}\ \emph {et~al.}(2026)\citenamefont
  {Tindall}, \citenamefont {Mello}, \citenamefont {Fishman}, \citenamefont
  {Stoudenmire},\ and\ \citenamefont {Sels}}]{Tindall2026}%
  \BibitemOpen
  \bibfield  {author} {\bibinfo {author} {\bibfnamefont {J.}~\bibnamefont
  {Tindall}}, \bibinfo {author} {\bibfnamefont {A.~F.}\ \bibnamefont {Mello}},
  \bibinfo {author} {\bibfnamefont {M.}~\bibnamefont {Fishman}}, \bibinfo
  {author} {\bibfnamefont {E.~M.}\ \bibnamefont {Stoudenmire}},\ and\ \bibinfo
  {author} {\bibfnamefont {D.}~\bibnamefont {Sels}},\ }\bibfield  {title}
  {\bibinfo {title} {Dynamics of disordered quantum systems with two- and
  three-dimensional tensor networks},\ }\href
  {https://doi.org/10.1126/science.adx2728} {\bibfield  {journal} {\bibinfo
  {journal} {Science}\ }\textbf {\bibinfo {volume} {392}},\ \bibinfo {pages}
  {868} (\bibinfo {year} {2026})}\BibitemShut {NoStop}%
\bibitem [{\citenamefont {Wiersema}(2026)}]{Wiersema2026}%
  \BibitemOpen
  \bibfield  {author} {\bibinfo {author} {\bibfnamefont {R.}~\bibnamefont
  {Wiersema}},\ }\href@noop {} {\bibinfo {title} {Numerical simulation of
  {D}-{W}ave's quantum advantage experiment with time-dependent variational
  {M}onte {C}arlo}} (\bibinfo {year} {2026}),\ \Eprint
  {https://arxiv.org/abs/2609.01719} {arXiv:2609.01719 [quant-ph]} \BibitemShut
  {NoStop}%
\bibitem [{\citenamefont {Zhang}\ \emph {et~al.}(2025)\citenamefont {Zhang},
  \citenamefont {Bayocboc},\ and\ \citenamefont {Dziarmaga}}]{Zhang2025}%
  \BibitemOpen
  \bibfield  {author} {\bibinfo {author} {\bibfnamefont {Y.}~\bibnamefont
  {Zhang}}, \bibinfo {author} {\bibfnamefont {F.~A.}\ \bibnamefont
  {Bayocboc}},\ and\ \bibinfo {author} {\bibfnamefont {J.}~\bibnamefont
  {Dziarmaga}},\ }\bibfield  {title} {\bibinfo {title} {{K}ibble-{Z}urek
  dynamical scaling hypothesis in the {G}oogle analog-digital quantum simulator
  of the {XX} model},\ }\href {https://doi.org/10.1103/s2fc-3d18} {\bibfield
  {journal} {\bibinfo  {journal} {Phys. Rev. B}\ }\textbf {\bibinfo {volume}
  {112}},\ \bibinfo {pages} {134420} (\bibinfo {year} {2025})}\BibitemShut
  {NoStop}%
\bibitem [{\citenamefont {Haghshenas}\ \emph {et~al.}(2026)\citenamefont
  {Haghshenas}, \citenamefont {Chertkov}, \citenamefont {Mills}, \citenamefont
  {Kadow}, \citenamefont {Lin}, \citenamefont {Chen}, \citenamefont {Cade},
  \citenamefont {Niesen}, \citenamefont {Begu{\v{s}}i{\'c}}, \citenamefont
  {Rudolph}, \citenamefont {Cirstoiu}, \citenamefont {H{\'e}mery},
  \citenamefont {Mc~Keever}, \citenamefont {Lubasch}, \citenamefont {Granet},
  \citenamefont {Baldwin}, \citenamefont {Bartolotta}, \citenamefont {Bohn},
  \citenamefont {Burau}, \citenamefont {Cline}, \citenamefont {DeCross},
  \citenamefont {Dreiling}, \citenamefont {Foltz}, \citenamefont {Francois},
  \citenamefont {Gaebler}, \citenamefont {Gilbreth}, \citenamefont {Gray},
  \citenamefont {Gresh}, \citenamefont {Hall}, \citenamefont {Hankin},
  \citenamefont {Hansen}, \citenamefont {Hewitt}, \citenamefont {Holliman},
  \citenamefont {Hutson}, \citenamefont {Iqbal}, \citenamefont {Kotibhaskar},
  \citenamefont {Lehman}, \citenamefont {Lucchetti}, \citenamefont {Madjarov},
  \citenamefont {Mayer}, \citenamefont {Milne}, \citenamefont {Moses},
  \citenamefont {Neyenhuis}, \citenamefont {Park}, \citenamefont {Perry},
  \citenamefont {Ponsioen}, \citenamefont {Schecter}, \citenamefont
  {Siegfried}, \citenamefont {Stephen}, \citenamefont {Tiemann}, \citenamefont
  {Urmey}, \citenamefont {Walker}, \citenamefont {Potter}, \citenamefont
  {Hayes}, \citenamefont {Chan}, \citenamefont {Pollmann}, \citenamefont
  {Knap}, \citenamefont {Dreyer},\ and\ \citenamefont
  {Foss-Feig}}]{Haghshenas2026}%
  \BibitemOpen
  \bibfield  {author} {\bibinfo {author} {\bibfnamefont {R.}~\bibnamefont
  {Haghshenas}}, \bibinfo {author} {\bibfnamefont {E.}~\bibnamefont
  {Chertkov}}, \bibinfo {author} {\bibfnamefont {M.}~\bibnamefont {Mills}},
  \bibinfo {author} {\bibfnamefont {W.}~\bibnamefont {Kadow}}, \bibinfo
  {author} {\bibfnamefont {S.-H.}\ \bibnamefont {Lin}}, \bibinfo {author}
  {\bibfnamefont {Y.~H.}\ \bibnamefont {Chen}}, \bibinfo {author}
  {\bibfnamefont {C.}~\bibnamefont {Cade}}, \bibinfo {author} {\bibfnamefont
  {I.}~\bibnamefont {Niesen}}, \bibinfo {author} {\bibfnamefont
  {T.}~\bibnamefont {Begu{\v{s}}i{\'c}}}, \bibinfo {author} {\bibfnamefont
  {M.~S.}\ \bibnamefont {Rudolph}}, \bibinfo {author} {\bibfnamefont
  {C.}~\bibnamefont {Cirstoiu}}, \bibinfo {author} {\bibfnamefont
  {K.}~\bibnamefont {H{\'e}mery}}, \bibinfo {author} {\bibfnamefont
  {C.}~\bibnamefont {Mc~Keever}}, \bibinfo {author} {\bibfnamefont
  {M.}~\bibnamefont {Lubasch}}, \bibinfo {author} {\bibfnamefont
  {E.}~\bibnamefont {Granet}}, \bibinfo {author} {\bibfnamefont {C.~H.}\
  \bibnamefont {Baldwin}}, \bibinfo {author} {\bibfnamefont {J.~P.}\
  \bibnamefont {Bartolotta}}, \bibinfo {author} {\bibfnamefont
  {M.}~\bibnamefont {Bohn}}, \bibinfo {author} {\bibfnamefont {J.~J.}\
  \bibnamefont {Burau}}, \bibinfo {author} {\bibfnamefont {J.}~\bibnamefont
  {Cline}}, \bibinfo {author} {\bibfnamefont {M.}~\bibnamefont {DeCross}},
  \bibinfo {author} {\bibfnamefont {J.~M.}\ \bibnamefont {Dreiling}}, \bibinfo
  {author} {\bibfnamefont {C.}~\bibnamefont {Foltz}}, \bibinfo {author}
  {\bibfnamefont {D.}~\bibnamefont {Francois}}, \bibinfo {author}
  {\bibfnamefont {J.~P.}\ \bibnamefont {Gaebler}}, \bibinfo {author}
  {\bibfnamefont {C.~N.}\ \bibnamefont {Gilbreth}}, \bibinfo {author}
  {\bibfnamefont {J.}~\bibnamefont {Gray}}, \bibinfo {author} {\bibfnamefont
  {D.}~\bibnamefont {Gresh}}, \bibinfo {author} {\bibfnamefont
  {A.}~\bibnamefont {Hall}}, \bibinfo {author} {\bibfnamefont {A.}~\bibnamefont
  {Hankin}}, \bibinfo {author} {\bibfnamefont {A.}~\bibnamefont {Hansen}},
  \bibinfo {author} {\bibfnamefont {N.}~\bibnamefont {Hewitt}}, \bibinfo
  {author} {\bibfnamefont {C.~A.}\ \bibnamefont {Holliman}}, \bibinfo {author}
  {\bibfnamefont {R.~B.}\ \bibnamefont {Hutson}}, \bibinfo {author}
  {\bibfnamefont {M.}~\bibnamefont {Iqbal}}, \bibinfo {author} {\bibfnamefont
  {N.}~\bibnamefont {Kotibhaskar}}, \bibinfo {author} {\bibfnamefont
  {E.}~\bibnamefont {Lehman}}, \bibinfo {author} {\bibfnamefont
  {D.}~\bibnamefont {Lucchetti}}, \bibinfo {author} {\bibfnamefont {I.~S.}\
  \bibnamefont {Madjarov}}, \bibinfo {author} {\bibfnamefont {K.}~\bibnamefont
  {Mayer}}, \bibinfo {author} {\bibfnamefont {A.~R.}\ \bibnamefont {Milne}},
  \bibinfo {author} {\bibfnamefont {S.~A.}\ \bibnamefont {Moses}}, \bibinfo
  {author} {\bibfnamefont {B.}~\bibnamefont {Neyenhuis}}, \bibinfo {author}
  {\bibfnamefont {G.}~\bibnamefont {Park}}, \bibinfo {author} {\bibfnamefont
  {A.~R.}\ \bibnamefont {Perry}}, \bibinfo {author} {\bibfnamefont
  {B.}~\bibnamefont {Ponsioen}}, \bibinfo {author} {\bibfnamefont
  {M.}~\bibnamefont {Schecter}}, \bibinfo {author} {\bibfnamefont {P.~E.}\
  \bibnamefont {Siegfried}}, \bibinfo {author} {\bibfnamefont {D.~T.}\
  \bibnamefont {Stephen}}, \bibinfo {author} {\bibfnamefont {B.~G.}\
  \bibnamefont {Tiemann}}, \bibinfo {author} {\bibfnamefont {M.~D.}\
  \bibnamefont {Urmey}}, \bibinfo {author} {\bibfnamefont {J.}~\bibnamefont
  {Walker}}, \bibinfo {author} {\bibfnamefont {A.~C.}\ \bibnamefont {Potter}},
  \bibinfo {author} {\bibfnamefont {D.}~\bibnamefont {Hayes}}, \bibinfo
  {author} {\bibfnamefont {G.~K.-L.}\ \bibnamefont {Chan}}, \bibinfo {author}
  {\bibfnamefont {F.}~\bibnamefont {Pollmann}}, \bibinfo {author}
  {\bibfnamefont {M.}~\bibnamefont {Knap}}, \bibinfo {author} {\bibfnamefont
  {H.}~\bibnamefont {Dreyer}},\ and\ \bibinfo {author} {\bibfnamefont
  {M.}~\bibnamefont {Foss-Feig}},\ }\bibfield  {title} {\bibinfo {title}
  {Digital quantum magnetism on a trapped-ion quantum computer},\ }\href
  {https://doi.org/10.1038/s41586-026-10445-3} {\bibfield  {journal} {\bibinfo
  {journal} {Nature}\ }\textbf {\bibinfo {volume} {653}},\ \bibinfo {pages}
  {56} (\bibinfo {year} {2026})}\BibitemShut {NoStop}%
\bibitem [{\citenamefont {Sinibaldi}\ \emph {et~al.}(2023)\citenamefont
  {Sinibaldi}, \citenamefont {Giuliani}, \citenamefont {Carleo},\ and\
  \citenamefont {Vicentini}}]{Sinibaldi2023}%
  \BibitemOpen
  \bibfield  {author} {\bibinfo {author} {\bibfnamefont {A.}~\bibnamefont
  {Sinibaldi}}, \bibinfo {author} {\bibfnamefont {C.}~\bibnamefont {Giuliani}},
  \bibinfo {author} {\bibfnamefont {G.}~\bibnamefont {Carleo}},\ and\ \bibinfo
  {author} {\bibfnamefont {F.}~\bibnamefont {Vicentini}},\ }\bibfield  {title}
  {\bibinfo {title} {Unbiasing time-dependent variational {M}onte {C}arlo by
  projected quantum evolution},\ }\href
  {https://doi.org/10.22331/q-2023-10-10-1131} {\bibfield  {journal} {\bibinfo
  {journal} {Quantum}\ }\textbf {\bibinfo {volume} {7}},\ \bibinfo {pages}
  {1131} (\bibinfo {year} {2023})}\BibitemShut {NoStop}%
\bibitem [{\citenamefont {Gravina}\ \emph {et~al.}(2025)\citenamefont
  {Gravina}, \citenamefont {Savona},\ and\ \citenamefont
  {Vicentini}}]{Gravina2025}%
  \BibitemOpen
  \bibfield  {author} {\bibinfo {author} {\bibfnamefont {L.}~\bibnamefont
  {Gravina}}, \bibinfo {author} {\bibfnamefont {V.}~\bibnamefont {Savona}},\
  and\ \bibinfo {author} {\bibfnamefont {F.}~\bibnamefont {Vicentini}},\
  }\bibfield  {title} {\bibinfo {title} {Neural projected quantum dynamics: a
  systematic study},\ }\href {https://doi.org/10.22331/q-2025-07-22-1803}
  {\bibfield  {journal} {\bibinfo  {journal} {Quantum}\ }\textbf {\bibinfo
  {volume} {9}},\ \bibinfo {pages} {1803} (\bibinfo {year} {2025})}\BibitemShut
  {NoStop}%
\bibitem [{\citenamefont {Hauschild}\ and\ \citenamefont
  {Pollmann}(2018)}]{HauschildPollmann2018}%
  \BibitemOpen
  \bibfield  {author} {\bibinfo {author} {\bibfnamefont {J.}~\bibnamefont
  {Hauschild}}\ and\ \bibinfo {author} {\bibfnamefont {F.}~\bibnamefont
  {Pollmann}},\ }\bibfield  {title} {\bibinfo {title} {Efficient numerical
  simulations with tensor networks: Tensor {N}etwork {P}ython ({T}e{NP}y)},\
  }\bibfield  {journal} {\bibinfo  {journal} {SciPost Phys. Lect. Notes}\
  }\textbf {\bibinfo {volume} {5}},\ \href
  {https://doi.org/10.21468/SciPostPhysLectNotes.5}
  {10.21468/SciPostPhysLectNotes.5} (\bibinfo {year} {2018})\BibitemShut
  {NoStop}%
\bibitem [{\citenamefont {Hauschild}\ \emph {et~al.}(2024)\citenamefont
  {Hauschild}, \citenamefont {Unfried}, \citenamefont {Anand}, \citenamefont
  {Andrews}, \citenamefont {Bintz}, \citenamefont {Borla}, \citenamefont
  {Divic}, \citenamefont {Drescher}, \citenamefont {Geiger}, \citenamefont
  {Hefel}, \citenamefont {H{\'e}mery}, \citenamefont {Kadow}, \citenamefont
  {Kemp}, \citenamefont {Kirchner}, \citenamefont {Liu}, \citenamefont
  {M{\"o}ller}, \citenamefont {Parker}, \citenamefont {Rader}, \citenamefont
  {Romen}, \citenamefont {Scalet}, \citenamefont {Schoonderwoerd},
  \citenamefont {Schulz}, \citenamefont {Soejima}, \citenamefont {Thoma},
  \citenamefont {Wu}, \citenamefont {Zechmann}, \citenamefont {Zweng},
  \citenamefont {Mong}, \citenamefont {Zaletel},\ and\ \citenamefont
  {Pollmann}}]{TeNPy2024}%
  \BibitemOpen
  \bibfield  {author} {\bibinfo {author} {\bibfnamefont {J.}~\bibnamefont
  {Hauschild}}, \bibinfo {author} {\bibfnamefont {J.}~\bibnamefont {Unfried}},
  \bibinfo {author} {\bibfnamefont {S.}~\bibnamefont {Anand}}, \bibinfo
  {author} {\bibfnamefont {B.}~\bibnamefont {Andrews}}, \bibinfo {author}
  {\bibfnamefont {M.}~\bibnamefont {Bintz}}, \bibinfo {author} {\bibfnamefont
  {U.}~\bibnamefont {Borla}}, \bibinfo {author} {\bibfnamefont
  {S.}~\bibnamefont {Divic}}, \bibinfo {author} {\bibfnamefont
  {M.}~\bibnamefont {Drescher}}, \bibinfo {author} {\bibfnamefont
  {J.}~\bibnamefont {Geiger}}, \bibinfo {author} {\bibfnamefont
  {M.}~\bibnamefont {Hefel}}, \bibinfo {author} {\bibfnamefont
  {K.}~\bibnamefont {H{\'e}mery}}, \bibinfo {author} {\bibfnamefont
  {W.}~\bibnamefont {Kadow}}, \bibinfo {author} {\bibfnamefont
  {J.}~\bibnamefont {Kemp}}, \bibinfo {author} {\bibfnamefont {N.}~\bibnamefont
  {Kirchner}}, \bibinfo {author} {\bibfnamefont {V.~S.}\ \bibnamefont {Liu}},
  \bibinfo {author} {\bibfnamefont {G.}~\bibnamefont {M{\"o}ller}}, \bibinfo
  {author} {\bibfnamefont {D.}~\bibnamefont {Parker}}, \bibinfo {author}
  {\bibfnamefont {M.}~\bibnamefont {Rader}}, \bibinfo {author} {\bibfnamefont
  {A.}~\bibnamefont {Romen}}, \bibinfo {author} {\bibfnamefont
  {S.}~\bibnamefont {Scalet}}, \bibinfo {author} {\bibfnamefont
  {L.}~\bibnamefont {Schoonderwoerd}}, \bibinfo {author} {\bibfnamefont
  {M.}~\bibnamefont {Schulz}}, \bibinfo {author} {\bibfnamefont
  {T.}~\bibnamefont {Soejima}}, \bibinfo {author} {\bibfnamefont
  {P.}~\bibnamefont {Thoma}}, \bibinfo {author} {\bibfnamefont
  {Y.}~\bibnamefont {Wu}}, \bibinfo {author} {\bibfnamefont {P.}~\bibnamefont
  {Zechmann}}, \bibinfo {author} {\bibfnamefont {L.}~\bibnamefont {Zweng}},
  \bibinfo {author} {\bibfnamefont {R.~S.~K.}\ \bibnamefont {Mong}}, \bibinfo
  {author} {\bibfnamefont {M.~P.}\ \bibnamefont {Zaletel}},\ and\ \bibinfo
  {author} {\bibfnamefont {F.}~\bibnamefont {Pollmann}},\ }\bibfield  {title}
  {\bibinfo {title} {Tensor network {P}ython ({T}e{NP}y) version 1},\ }\href
  {https://doi.org/10.21468/SciPostPhysCodeb.41} {\bibfield  {journal}
  {\bibinfo  {journal} {SciPost Phys. Codebases}\ ,\ \bibinfo {pages} {41}}
  (\bibinfo {year} {2024})}\BibitemShut {NoStop}%
\bibitem [{\citenamefont {Bradbury}\ \emph {et~al.}(2018)\citenamefont
  {Bradbury}, \citenamefont {Frostig}, \citenamefont {Hawkins}, \citenamefont
  {Johnson}, \citenamefont {Leary}, \citenamefont {Maclaurin}, \citenamefont
  {Necula}, \citenamefont {Paszke}, \citenamefont {VanderPlas}, \citenamefont
  {Wanderman-Milne},\ and\ \citenamefont {Zhang}}]{JAX2018}%
  \BibitemOpen
  \bibfield  {author} {\bibinfo {author} {\bibfnamefont {J.}~\bibnamefont
  {Bradbury}}, \bibinfo {author} {\bibfnamefont {R.}~\bibnamefont {Frostig}},
  \bibinfo {author} {\bibfnamefont {P.}~\bibnamefont {Hawkins}}, \bibinfo
  {author} {\bibfnamefont {M.~J.}\ \bibnamefont {Johnson}}, \bibinfo {author}
  {\bibfnamefont {C.}~\bibnamefont {Leary}}, \bibinfo {author} {\bibfnamefont
  {D.}~\bibnamefont {Maclaurin}}, \bibinfo {author} {\bibfnamefont
  {G.}~\bibnamefont {Necula}}, \bibinfo {author} {\bibfnamefont
  {A.}~\bibnamefont {Paszke}}, \bibinfo {author} {\bibfnamefont
  {J.}~\bibnamefont {VanderPlas}}, \bibinfo {author} {\bibfnamefont
  {S.}~\bibnamefont {Wanderman-Milne}},\ and\ \bibinfo {author} {\bibfnamefont
  {Q.}~\bibnamefont {Zhang}},\ }\href@noop {} {\bibinfo {title} {{JAX}:
  composable transformations of {P}ython+{N}um{P}y programs}},\ \bibinfo
  {howpublished} {\url{https://github.com/jax-ml/jax}} (\bibinfo {year}
  {2018})\BibitemShut {NoStop}%
\bibitem [{\citenamefont {Wu}(2026)}]{Data2026}%
  \BibitemOpen
  \bibfield  {author} {\bibinfo {author} {\bibfnamefont {Y.}~\bibnamefont
  {Wu}},\ }\href@noop {} {\bibinfo {title} {Data repository for ``{{E}}fficient
  simulation of long-range systems: ground state and real-time dynamics''}},\
  \bibinfo {howpublished}
  {\url{https://github.com/yantaow/open_data/tree/main/wu2026efficient}}
  (\bibinfo {year} {2026})\BibitemShut {NoStop}%
\end{thebibliography}%
\onecolumngrid
\setcounter{equation}{0}
\newpage
\renewcommand{\thesection}{S-\arabic{section}} \renewcommand{\theequation}{S%
\arabic{equation}} \setcounter{equation}{0} \renewcommand{\thefigure}{S%
\arabic{figure}} \setcounter{figure}{0}\setcounter{section}{0}
\renewcommand{\thetable}{S\arabic{table}}\setcounter{table}{0}
\renewcommand{\theHequation}{S\arabic{equation}}
\renewcommand{\theHfigure}{S\arabic{figure}}
\renewcommand{\theHtable}{S\arabic{table}}
\renewcommand{\theHsection}{S\arabic{section}}
\centerline{\textbf{Supplemental Material}}
\maketitle
\newcommand{\SMtocoff}{\let\SMcontentsline\contentsline\renewcommand{\contentsline}[4]{}}
\newcommand{\SMtocon}{\let\contentsline\SMcontentsline}
\addtocontents{toc}{\protect\SMtocon}
\tableofcontents

\section{Adaptive-basis sampling: general formulation and cost}
\label{sec:sm_general}

\textit{Rotated tensor network states.}
Write the TNS in the reference ($z$) basis as $\ket{\psi_{\vec\theta}}=\sum_{\vec s}\mathcal C\big(\{A_i^{s_i}\}\big)\ket{\vec s}$, where $\mathcal C$ denotes the contraction of the virtual indices and $\vec\theta$ collects the tensor elements.
For a product basis $\ket{\vec s_b}=\bigotimes_i U_b^\dagger\ket{s_i}$ with a single-site unitary $U_b$,
\begin{equation}
  \braket{\vec s_b|\psi_{\vec\theta}}=\mathcal C\big(\{\tilde A_i^{s_i}\}\big),\qquad
  \tilde A_i^{s}=\sum_{s'}\braket{s|U_b|s'}A_i^{s'} .
\end{equation}
The rotated network has the same bond dimension, and forming it costs $O(Nd^2D^4)$ for physical dimension $d$ and a square-lattice TNS of bond dimension $D$, which is negligible.
The map $\vec\theta\mapsto\tilde{\vec\theta}=U_b\vec\theta$ is linear and unitary, acting block-diagonally on the physical index of each tensor.
For a generic neural quantum state, by contrast, $\braket{\vec s_b|\psi}=\sum_{\vec z}\braket{\vec s_b|\vec z}\psi(\vec z)$ involves $2^N$ terms.

\textit{Equations of motion.}
Let $H=\sum_b H_b$.
In the parameters $\tilde{\vec\theta}_b=U_b\vec\theta$, the quantum geometric tensor and force estimated from samples $\vec s_b\sim|\braket{\vec s_b|\psi_{\vec\theta}}|^2$ are $S^{(b)}=U_bSU_b^\dagger$ and $\vec F^{(b)}=U_b\vec F_b$, where $S$ is the geometric tensor in the reference parameters and $\vec F_b$ is the force of $H_b$.
Because $\mathbf{F} = \sum_b \mathbf{F}_b$,
\begin{equation}
    \dot{\boldsymbol{\theta}} = -\gamma S^{-1} \sum_b \mathbf{F}_b = \sum_b U_b^\dagger \bigl(-\gamma S^{(b)-1} \mathbf{F}^{(b)}\bigr) \equiv \sum_b U_b^\dagger \dot{\boldsymbol{\theta}}_b,
\end{equation}
which yields Eq.~\eqref{eq:sum} of the main text for both imaginary-time ($\gamma = 1$) and real-time ($\gamma = i$) evolution.
In practice, each chain provides its own estimate of $S^{(b)}$, and the corresponding linear system is solved independently with the same regularization; in the infinite-sample limit, the result coincides with a single solve with the total force.

\textit{Assignment of terms.}
For the general two-body Hamiltonian, Eq.~\eqref{eq:general} of the main text, we assign every term whose first Pauli operator is $\sigma^a$ to the basis $b=a$.
In that basis, the local energy is
\begin{equation}
  H_{b,\loc}(\vec s)=\sum_{i<j}J^{bb}_{ij}s_is_j+\sum_{i<j}\sum_{a'\neq b}J^{ba'}_{ij}s_i\,R^{a'}_j(\vec s)+\sum_i\sum_{a}h_i^{a}R_i^{a}(\vec s),
  \qquad
  R^{a}_j(\vec s)=\frac{\braket{\vec s|\sigma^{a}_j|\psi}}{\braket{\vec s|\psi}},
  \label{eq:sm_Hloc}
\end{equation}
where $R^{b}_j=s_j$ and, for $a\neq b$, $R^a_j$ is a phase times the amplitude ratio for flipping spin $j$; the field terms are assigned to any convenient basis.
The $N$ single-site ratios $R^a_j$ are obtained from the left, right, upper, and lower environments that are cached during a sampling sweep, at a total cost of $O(N)$ local contractions, the same as for the log-derivatives $O_\mu$.
The remaining pair sums in Eq.~\eqref{eq:sm_Hloc} are scalar operations, $O(N^2)$ per sample, or $O(N\log N)$ by fast Fourier transform for translation-invariant couplings.
For the XY Hamiltonians of the main text, $J^{aa'}_{ij}=0$ for $a\neq a'$ and two bases ($x$ and $y$) suffice; the staggered $\sigma^z$ field of the Rydberg protocol is placed in the $y$ chain, where it requires only the single-site ratios $R^z_j$.
A long-range $\sigma^z_i\sigma^z_j$ term, as in a dipolar Heisenberg or XXZ model, requires a third chain in the $z$ basis.

\textit{Comparison with the fixed basis.}
In the $z$ basis, $\sigma_i^x\sigma_j^x$ maps $\vec z$ to the configuration $\vec z^{ij}$ with spins $i$ and $j$ flipped, and its local energy requires the ratio $\psi(\vec z^{ij})/\psi(\vec z)$.
Evaluating this ratio requires recontracting the entire single-layer TNS for the amplitude at $O(N)$ cost [reusing cached boundary environments can reduce this by a constant factor of $2$--$3$, but the $O(N)$ scaling remains].
With $O(N^2)$ such pairs, the fixed-basis local energy costs $O(N^3)$ contractions per sample, compared with $O(N)$ for sampling the configuration itself.

\section{Ground states: parameters and extrapolation}
\label{sec:sm_gs}

\textit{Optimization and measurement.}
The ground states of $H_\alpha$ in Fig.~\ref{fig:alpha} are obtained by imaginary-time evolution with stochastic reconfiguration on open $L\times L$ arrays, imposing $C_{2v}$ symmetry.
For all $\alpha$ and $L$, we use TNS bond dimension $D=4$ with boundary-MPS bond dimension $D'=12$ for contraction, $81920$ samples per basis and step ($N_{\mathrm{chain}}=8192$ parallel Markov chains with $N_{\mathrm{s/c}}=10$ sequential samples per chain), an imaginary-time step $\delta\tau=0.02$, and a diagonal shift of $10^{-3}$ for $S$.
Final observables are evaluated using $8192$ independent chains per basis ($200$ burn-in and $400$ measurement sweeps), with error bars estimated by the jackknife resampling across chains.

\begin{figure}[t]
\centering
\includegraphics[width=\linewidth]{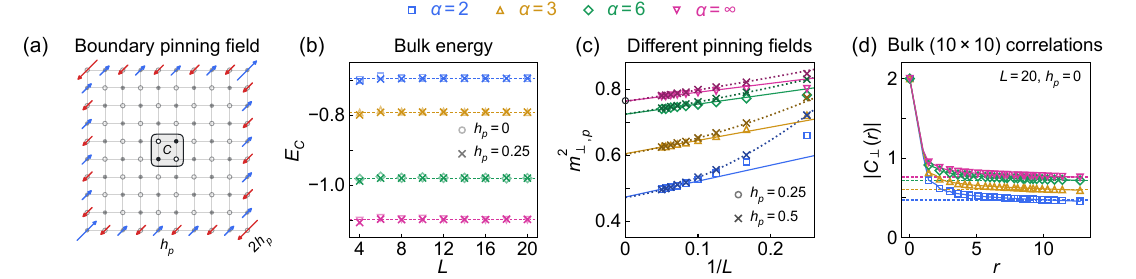}
\caption{Pinning-field method.
(a)~Schematic of the boundary pinning field on a $10\times10$ array and the central $2\times2$ block $C$.
(b)~Bulk energy per site $E_C$ (measured on block $C$) versus $L$ without ($h_p=0$, open symbols) and with ($h_p=0.25$, crosses) the pinning field.
Dashed lines indicate the extrapolated total energy per site $E/N$.
(c)~Pinned staggered magnetization $m_{\perp,p}^2$ versus $1/L$ for $h_p=0.25$ (open symbols) and $h_p=0.5$ (crosses). Solid lines are weighted linear fits in $1/L$ over $L=14$--$20$ for
$h_p=0.25$ (as in the main text), and dotted lines are weighted quadratic fits over $L\ge8$ for $h_p=0.5$.
The black open circle at $1/L=0$ marks the $\alpha=\infty$ SSE result~\cite{Sandvik1999}.
In (b) and (c), darker color shades correspond to larger $h_p$.
(d)~Unpinned correlation $|C_\perp(r)|$ on the $20\times20$ array ($h_p=0$), averaged over site pairs within the central $10\times10$ block. Dashed lines show the $h_p=0.25$ extrapolated values from~(c).}
\label{fig:sm_pinning}
\end{figure}

\textit{Pinning-field method.}
On open arrays, boundary effects suppress the two-point correlation $|C_\perp(r)|$ at large distances [$r\gtrsim L/2$ in Fig.~\ref{fig:alpha}(c)], making it difficult to extract the in-plane AFM order parameter directly.
We therefore employ the pinning-field method~\cite{WhiteChernyshev2007,AssaadHerbut2013} by adding a staggered in-plane field along the boundary,
\begin{equation}
    H_p=h_p\sum_i n_i\,\epsilon_i\,\frac{\sigma_i^x+\sigma_i^y}{\sqrt2},
\end{equation}
where $n_i$ is the number of missing nearest-neighbor bonds of site $i$ ($n_i=1$ on edges, $2$ at corners, and $0$ in the bulk) and we set $h_p=0.25$.
Acting only on the $O(L)$ boundary sites, $H_p$ serves as a symmetry-breaking boundary condition~\cite{WhiteChernyshev2007} that fixes the N\'eel direction without altering the bulk property.
The order parameter is then extracted from the staggered moment $m_{\perp,p}^2(L;h_p)$ [Eq.~\eqref{eq:mpin}] induced at the central sites (a distance $\sim L/2$ from the edges), which converges in the thermodynamic limit to the spontaneous value
\begin{equation}
    m^2=\lim_{L\to\infty}m_{\perp,p}^2(L;h_p), \label{eq:sm_pin_limit}
\end{equation}
requiring only a finite-size extrapolation at fixed $h_p$ without a subsequent $h_p\to0$ limit~\cite{AssaadHerbut2013}.
The pinned states are optimized using the same settings as above.
As benchmarked in Fig.~\ref{fig:sm_pinning}, the pinning field leaves the bulk energy virtually unchanged and consistent with the $1/L$-extrapolated total energy per site [Fig.~\ref{fig:alpha}(b), Fig.~\ref{fig:sm_pinning}(b)].
Moreover, the extrapolated $m_{\perp,p}^2$ values for $h_p=0.25$ and $0.5$ agree closely [Fig.~\ref{fig:sm_pinning}(c)] and match the long-distance plateau of the unpinned bulk correlation $|C_\perp(r)|$ [Fig.~\ref{fig:sm_pinning}(d)].

\textit{Extrapolation.}
The energy per site $e(L)=E(L)/N$ [Fig.~\ref{fig:alpha}(b)] and staggered magnetization $m_{\perp,p}^2(L)$ [Fig.~\ref{fig:alpha}(c), inset] are extrapolated to the thermodynamic limit ($L\to\infty$) by fitting the results for $L=14, 16, 18,$ and $20$ to the linear form
\begin{equation}
    y(L)=y_\infty+\frac{b}{L}. 
    \label{eq:sm_fit}
\end{equation}
The fit minimizes $\chi^2=\sum_L[y(L)-y_\infty-b/L]^2/\sigma_L^2$, where $\sigma_L$ is the jackknife standard error at size $L$.
Whenever $\chi^2/\text{dof}>1$ (with $\text{dof}=2$ degrees of freedom), the uncertainty of $y_\infty$ is rescaled by $\sqrt{\chi^2/\text{dof}}$.
The extrapolated values are summarized in Table~\ref{tab:extrap}, and the finite-size data entering the fits are listed in Table~\ref{tab:sm_finite}.

\section{Uncertainty relation}
\label{sec:sm_canting}
We define the uniform in-plane, staggered in-plane, and staggered out-of-plane moments as
\begin{equation}
    M_a=\frac{1}{\sqrt N}\sum_i\sigma_i^a,\qquad N_a=\frac1N\sum_i\epsilon_i\sigma_i^a,\qquad Z=\frac{1}{\sqrt N}\sum_i\epsilon_i\sigma_i^z,
\end{equation}
with $\epsilon_i=(-1)^{x_i+y_i}$ and $a=x,y,z$.
Since $[\sigma_i^a,\sigma_j^b]=2i\delta_{ij}\varepsilon_{abc}\sigma_i^c$, we have $[M_x,Z]=-2iN_y$ and $[M_y,Z]=2iN_x$.
Applying the Robertson uncertainty relation~\cite{Robertson1929}
\begin{equation}
    \braket{A^2}\braket{B^2}\ge\Delta A^2\Delta B^2\ge\tfrac14\big|\braket{[A,B]}\big|^2
\end{equation}
to $(M_x,Z)$ and $(M_y,Z)$ and summing the two inequalities yields
\begin{equation}
    \braket{M_\perp^2}\braket{Z^2}\ \ge\ \braket{N_x}^2+\braket{N_y}^2,
    \label{eq:sm_rob_raw}
\end{equation}
where $M_\perp^2=M_x^2+M_y^2$.

On a finite system without symmetry breaking, $\braket{N_x}=\braket{N_y}=0$, so Eq.~\eqref{eq:sm_rob_raw} reduces to the trivial bound $\braket{M_\perp^2}\braket{Z^2}\ge0$.
In the thermodynamic limit ($L\to\infty$), however, spontaneous symmetry breaking yields a nonzero order parameter $\braket{N_x}^2+\braket{N_y}^2\to m_\perp^2$, leading to Eq.~\eqref{eq:robertson}:
\begin{equation}
    \eta=\frac{\braket{M_\perp^2}\braket{Z^2}}{m_\perp^2}\ \ge\ 1\qquad(L\to\infty).
\end{equation}
In the main text, these $U(1)$-invariant quantities are evaluated in the symmetric state, where $m_\perp^2=\braket{N_x^2+N_y^2}$ approaches the spontaneous value only as $L\to\infty$; consequently, $\eta$ can fall below $1$ at small sizes and satisfies $\eta\ge1$ only for sufficiently large $L$.

\begin{table}[t]
\centering
\renewcommand{\arraystretch}{1.2}
\setlength{\tabcolsep}{4pt}
\caption{Finite-size ground-state energy per site $E/N$ and staggered magnetization $m_{\perp,p}^2$ on open $L\times L$ arrays.}
\label{tab:sm_finite}
\begin{tabular}{c|ccccc|cccc}
\hline\hline
 & \multicolumn{5}{c|}{$E/N$} & \multicolumn{4}{c}{$m_{\perp,p}^2$}\\
$L\backslash\alpha$ & $1$ & $2$ & $3$ & $6$ & $\infty$ & $2$ & $3$ & $6$ & $\infty$\\
\hline
4 & $-0.57142(5)$ & $-0.6383(1)$ & $-0.6975(2)$ & $-0.8089(3)$ & $-0.8756(4)$ & $0.6597(8)$ & $0.7210(8)$ & $0.7819(9)$ & $0.8014(9)$\\
6 & $-0.57825(4)$ & $-0.65534(9)$ & $-0.7265(2)$ & $-0.8637(3)$ & $-0.9459(5)$ & $0.5793(8)$ & $0.6785(8)$ & $0.7719(9)$ & $0.8002(9)$\\
8 & $-0.58161(3)$ & $-0.66455(9)$ & $-0.7421(2)$ & $-0.8919(4)$ & $-0.9829(5)$ & $0.5454(7)$ & $0.6574(8)$ & $0.7606(9)$ & $0.7941(9)$\\
10 & $-0.58364(2)$ & $-0.6700(1)$ & $-0.7515(2)$ & $-0.9081(5)$ & $-1.0040(6)$ & $0.5259(7)$ & $0.6449(8)$ & $0.7531(9)$ & $0.789(1)$\\
12 & $-0.58488(2)$ & $-0.6745(1)$ & $-0.7580(2)$ & $-0.9205(5)$ & $-1.0199(8)$ & $0.5152(8)$ & $0.6378(8)$ & $0.7486(9)$ & $0.785(1)$\\
14 & $-0.58595(2)$ & $-0.6772(1)$ & $-0.7632(3)$ & $-0.9288(7)$ & $-1.0324(9)$ & $0.5074(7)$ & $0.6335(8)$ & $0.7458(9)$ & $0.783(1)$\\
16 & $-0.58667(2)$ & $-0.6797(1)$ & $-0.7662(3)$ & $-0.9349(9)$ & $-1.041(1)$ & $0.5036(8)$ & $0.6293(8)$ & $0.7439(9)$ & $0.781(1)$\\
18 & $-0.58722(1)$ & $-0.6813(1)$ & $-0.7692(3)$ & $-0.9410(7)$ & $-1.047(1)$ & $0.4999(8)$ & $0.6263(8)$ & $0.7400(9)$ & $0.778(1)$\\
20 & $-0.58765(1)$ & $-0.6824(1)$ & $-0.7721(2)$ & $-0.9446(6)$ & $-1.052(1)$ & $0.4971(8)$ & $0.6251(9)$ & $0.7398(9)$ & $0.778(1)$\\
\hline\hline
\end{tabular}
\end{table}

\section{Equivalence of the antiferromagnetic protocol and ground-state preparation}
\label{sec:sm_equiv}
Because $H(t)$ in Eq.~\eqref{eq:Ht} of the main text is a real matrix in the $z$ basis and the N\'eel initial state has real amplitudes, complex conjugation of the Schr\"odinger equation, $i\partial_t\ket{\psi}=H\ket{\psi}$, gives $i\partial_t\ket{\psi^*}=-H\ket{\psi^*}$.
The state evolved with $H(t)$ is therefore the complex conjugate of the state evolved with
\begin{equation}
  -H(t)=H_{\alpha=3}+|\delta(t)|\sum_{i\in B}\frac{1+\sigma_i^z}{2}\qquad(\delta_0<0),
\end{equation}
and expectation values of real operators, including all $\sigma^a_i\sigma^a_j$ correlations, coincide.
The N\'eel state is the ground state of $-H(0)$ in the $S^z_\text{tot}=0$ sector for large $|\delta_0|$, so the $\delta_0<0$ protocol adiabatically prepares the ground state of $H_{\alpha=3}$.

\section{Unit conversion in the Rydberg dynamics}
Using the notation of Ref.~\cite{Chen2023}, the Rydberg dynamics is prescribed by
\begin{align}
  H_{XY} &= -\frac{J}{2} \sum_{i<j} \left(\frac{a}{r_{ij}}\right)^3 (\sigma_i^x \sigma_j^x + \sigma_i^y\sigma_j^y),
\\
H_Z(t) &= \hbar \delta(t) \sum_{i \in B} n_i, \qquad n_i = \frac{1+\sigma_i^z}{2} = \mm{1}{0}{0}{0},
\\
H(t) &= H_{XY} + H_Z(t),
\end{align}
where $\sigma^a_i$ are Pauli matrices and $\delta(t) = \delta_0 e^{-t/\tau}$, with
$J/h = 0.77~\text{MHz}$, $\vert\delta_0\vert = 2\pi \times 15~\text{MHz}$, $\tau = 0.3~\mu\text{s}$, and total ramp time $T = 2~\mu\text{s}$.
In dimensionless units:
\begin{enumerate}
  \item $e^{-iHt/\hbar} = e^{-i\tilde{H} \tilde{t}}$. We take $\tilde{t} \equiv Jt/\hbar$ and $\tilde{r} = r/a$, so that
    \begin{equation*}
    \tilde{H} = \frac{H}{J} =  -\frac{1}{2} \sum_{i<j} \frac{1}{\tilde{r}^3_{ij}} (\sigma_i^x\sigma_j^x + \sigma_i^y\sigma_j^y) + \tilde{\delta}_0e^{-\tilde{t}/\tilde{\tau}} \sum_{i\in B} \frac{1+\sigma_i^z}{2}.
    \end{equation*}
    $\tilde{t} = 1$ corresponds to $t = \hbar/J = 0.2067~\mu\text{s}$.
  \item $\tilde{\tau} = \tau J/\hbar = 0.3~\mu\text{s} \times 2\pi \times 0.77~\text{MHz} = 1.45$.
  \item $\tilde{\delta}_0 = 2\pi \times 15~\text{MHz} \times \hbar/J = 15~\text{MHz}/0.77~\text{MHz} = 19.48$.
  \item $\tilde{T} = J T/\hbar = 9.68$.
\end{enumerate}
The initial state is prepared as the ground state in the limit $\tilde{\delta}_0 = \infty$.
Dropping the tildes gives the dimensionless variables used in the main text.

\section{Real-time dynamics: parameters and convergence}

Table~\ref{tab:hyper_parameters} lists the real-time simulation parameters.

\begin{table}[t]
\centering
\renewcommand{\arraystretch}{1.3}
\setlength{\tabcolsep}{8pt}
\caption{Simulation parameters for the real-time dynamics. $L$, side length. $D$, TNS bond dimension; the corresponding boundary MPS bond dimension is set to $D^\prime=4D$. $N_p$, number of independent variational parameters. $N_\text{chain}$, number of Markov chains, which also serves as the computational batch size. $N_\text{s/c}$, sequential samples per chain. $t_\text{wall}$, wall time per tVMC step in seconds, measured on a single GPU with 141 GB memory and 34 TFLOPS of FP64 performance. All simulations use a fourth-order Runge--Kutta integrator with time step $dt=0.02$.}
\begin{tabular}{cccccc}
\hline\hline
$L$ & $D$ & $N_p$ & $N_\text{chain}$ & $N_\text{s/c}$ & $t_\text{wall}/\text{s}$\\
\hline
4  & 4 & 840   & 16384 & 5  & 19\\
4  & 5 & 1860  & 16384 & 5  & 55\\
10 & 4 & 9384  & 8192  & 10 & 368\\
10 & 4 & 9384  & 8192  & 20 & 747\\
10 & 5 & 22260 & 8192  & 10 & 1158\\
10 & 5 & 22260 & 8192  & 20 & 3260\\
10 & 5 & 22260 & 8192  & 30 & 4309\\
10 & 5 & 22260 & 8192  & 40 & 5302\\
\hline\hline
\end{tabular}
\label{tab:hyper_parameters}
\end{table}

\begin{figure}[h]
\centering
\includegraphics[width=0.35\linewidth]{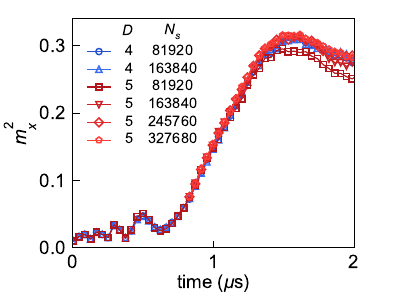}
\caption{Convergence of the squared staggered magnetization $m_x^2$ on a $10\times10$ array with respect to the TNS bond dimension $D$ and sample size $N_s$ (Table~\ref{tab:hyper_parameters}).}
\label{fig:conv}
\end{figure}

Figure~\ref{fig:conv} checks the convergence of the $10\times10$ dynamics with respect to the TNS bond dimension $D$ and sample size $N_s$. 
We focus on the staggered magnetization $m_x^2$, which is more sensitive than the energy (Fig.~\ref{fig:4x4}). 
Because a larger $D$ requires more samples, the $D=5$ curve with $N_s=81920$ deviates slightly at late times, whereas for $N_s\ge163840$ the $D=4$ and $D=5$ results become nearly identical, confirming convergence in both $D$ and $N_s$.

\end{document}